\documentclass[conf]{new-aiaa}
\usepackage[utf8]{inputenc}

\usepackage{graphicx}
\usepackage{amsmath}
\usepackage[version=4]{mhchem}
\usepackage{siunitx}
\usepackage{longtable,tabularx}
\newcommand{\R}{\mathbb{R}}
\newcommand{\rd}{\mathrm{d}}

\title{Modelling visibility and surface deformation in particle-fluid flow fields generated by helicopter rotors}

\author{S. Langdon\footnote{Professor of Mathematics, Department of Mathematics, Brunel University of London}}
\affil{Department of Mathematics, Brunel University of London,
Uxbridge, UB8 3PH, United Kingdom}
\author{D. J. Needham\footnote{Professor of Mathematics, School of Mathematics, University of Birmingham}}
\affil{School of Mathematics, University of Birmingham, Birmingham, B15 2TT, United Kingdom}

\begin{document}

\maketitle

\begin{abstract}
As a helicopter descends towards a bed of sand, a high velocity particle laden cloud can form around the helicopter body, a phenomenon known as “brownout”, and a consequence of which can potentially be a significant deterioration in visibility for the helicopter pilot.  Here we consider a recently developed physically based rational mathematical model for the generation of wind-driven particle flow fields from otherwise static particle beds, one application of which is the scenario considered here.  We introduce a directional opacity measure, defined for each observation angle from the helicopter cockpit, and show how visibility may vary in the model as certain parameters are varied.  In particular, we demonstrate a counterintuitive result suggesting that, with specific yet potentially realistic parameter choices, pilot visibility may be improved in some viewing directions if the helicopter were hovering at a lower altitude.  We also calculate the associated deformation of the upper surface of the particle bed, and show how certain surface deformation features may be sensitive to variation of key parameters.
\end{abstract}

\section{Nomenclature}

{\renewcommand\arraystretch{1.0}
\noindent\begin{longtable*}{@{}l @{\quad=\quad} l@{}}
$z$  & dimensionless distance measured vertically upwards, with $z=0$ the undisturbed level of the particle bed \\
$R$ & dimensionless distance measured horizontally, with $R=0$ the horizontal location of the helicopter \\
$\vartheta$ & azimuthal angle, with radial symmetry in $\vartheta$ assumed \\
$E$ & voidage field (representing the volume of fluid per unit spatial volume) \\
$E_s$ & dimensionless parameter, packing voidage of the sand particles \\
$\omega$ & domain on which the boundary value problem is posed \\
$z_d$ & dimensionless parameter, ratio of the hovering height of the helicopter rotor to the rotor blade radius\\
$\Delta$ & dimensionless parameter measuring the ratio of the helicopter rotor core radius to the rotor blade radius\\
$z_d^+$, $z_d^-$ & upper and lower vertical boundaries of cylindrical shell with dipole at its centre \\
$g$ & boundary data \\
$\alpha$ & dimensionless parameter, ratio of the particle collisional pressure to the drag force in the particle flow phase \\
$\gamma$ & dimensionless parameter, ratio of lift force per unit volume to  gravity force per unit volume in interfacial layer\\
$\Omega$ & dimensionless parameter, ratio of hovering helicopter induced swirl velocity to induced downwash velocity\\
$\theta$ & observation angle\\
$Op(\theta)$ & directional opacity measure\\
$\cal{C}(\theta)$ & integration contour for directional opacity measure\\
$\mathcal{B}$ & brownout set, consisting of those angles where total brownout is recorded\\
$t$ & dimensionless time\\
$\xi(R,t)$ & surface deformation, time and space dependent\\
$\chi(R)$ & surface deformation, spatial component\\
$h$ & mesh width for numerical approximation scheme\\
$\bar{\beta}(E)$ & decreasing function of $E$, representing the effect of neighbouring particles on the Stokes drag
\end{longtable*}}

\section{Introduction}
\label{sec:intro}
\lettrine{I}{n} this paper, we consider the scenario in which a helicopter is descending towards a bed of sand.  The local interaction between the down-draft and swirling flow, generated by the helicopter rotor, and the upper surface of the otherwise static sand bed, entrains sand particles from a thin interfacial layer into the air, which we consider as a fluidized region in which the particles lifted from the surface of the static particle bed are in suspension in the fluid, forming a fully developed two-phase flow.  These sand particles then flow in the form  of a high velocity particle laden cloud around the helicopter body, which we refer to as the \emph{helicopter cloud}, and a consequence of which is generally a significant deterioration in visibility for the helicopter pilot - when this becomes too severe, it is referred to as \emph{brownout}.

This problem has received much attention in the aero-engineering literature.  A common approach (see, e.g., \cite{GLG,PoMiPaPeVe20,TaGaBaLiZhHu21,TGZWCW,TaYoHeYuWa22,LSXW,LXJHL})
is to treat each particle within the fluid flow as an individual entity, with each particle tracked as it moves within the fluid flow according to its own dynamical equation of motion under the action of the locally induced fluid interaction forces and gravity, but with no consideration of the entrainment or detrainment of particles into the particle laden flow via local interaction of the flow with the upper surface of the otherwise static particle bed.  This is a severe drawback for this modelling approach, as it is this interaction process that is the fundamental and key process in initiating and driving the whole phenomenon of the particle cloud.  Moreover, given that particle clouds in the fluidized region generally have a high particle concentration, the approach of treating each particle in the fluidized region as an individual entity may be computationally inefficient in comparison to a more natural two-phase flow approach.

An alternative approach is to introduce a continuum particle density field, measuring particle volume per unit spatial volume of the two-phase flow and satisfying an suitable advection-diffusion partial differential equation (PDE) in the fluidized region.  This approach was developed by \cite{PhKiBr11}, who addressed the entrainment and detrainment of particles from the static particle bed via the introduction of a particle mass source term into the advection-diffusion PDE, localised in space, so as to act only in a thin neighbourhood of the interface between the fluidized region and the static particle bed, and designed to represent the localised input/output of particles from the static bed into the fluidized region.  However, the nature of the source term in \cite{PhKiBr11} is purely phenomenological, empirically based on the very particular situation under immediate consideration, and must be recalibrated in every specific example.  This limits the ability of this approach to capture, in general, a decent representation of the key rational mechanism of entrainment/detrainment at the interface of the fluidized region and the static bed, a point fully addressed in \cite{PhKiBr11}.  This specific issue was considered in~\cite{NL24}, where the authors developed
from fundamental first principles a physically based rational (consistent with the fundamental Newtonian laws of mechanics) mathematical model for the generation of wind-driven particle flow fields from otherwise static particle beds, leading to a natural macroscopic boundary condition on void fraction to be applied at the interface of the fluidized region and the static bed, and hence closing the continuum scale problem two-phase flow in the fluidized region.

Results in~\cite{NL24} demonstrated how the voidage field, representing the volume of fluid per unit spatial volume, could be computed through solution of a nonlinear elliptic boundary value problem.  We begin here, in~\S\ref{sec:model} by stating this boundary value problem and summarising its key features.  We then proceed, in~\S\ref{sec:opacity}, by introducing a new ``opacity'' function, that measures the visibility from the helicopter in all radial directions, and show how this can be easily computed from the solution of the nonlinear boundary value problem (equations (\ref{eqn:h34})--(\ref{eqn:h38}) below).  We present results in~\S\ref{sec:opacity} demonstrating how the opacity, and hence the visibility, vary with key parameter choices, and present a set of results, most of which align well with our intuition, but some of which demonstrate potentially counterintuitive phenomena, such as that the visibility can improve if the helicopter is hovering closer to the particle bed, under certain parameter choices.  In~\S\ref{sec:deformation}, we introduce a new ``surface deformation'' function, derived directly from the theory presented in ~\cite{NL24}, that measures how the initially flat surface is deformed by the hovering helicopter, and again show how this can be easily computed from the solution of (\ref{eqn:h34})--(\ref{eqn:h38}). Finally, in \S\ref{sec:conc}, we present some conclusions.

\section{The nonlinear elliptic boundary value problem}
\label{sec:model}

For a helicopter hovering steadily, with rotor blades rotating in a horizontal plane, in otherwise still air above a sand bed, which has undisturbed level at $z=0$, the steady problem for the associated voidage field in the fluid flow region can be stated in cylindrical polar coordinates $(R,\vartheta,z)$ (with $z$ measuring dimensionless distance vertically upwards), with radial symmetry in $\vartheta$, as the elliptic boundary value problem:
\begin{eqnarray}
  & E\left(E_{RR} + \frac{1}{R} E_R + E_{zz} \right) - \bar{a}(R,z) E_R - \bar{b}(R,z) E_z = 0, \quad (R,z)\in \omega,& \label{eqn:h34} \\
  & E=g(R) \mbox{ on } z=0, \quad R\geq 0, & \label{eqn:h35} \\
  & E_R = 0 \mbox{ on } \displaystyle{\left\{ \begin{array}{ll} R=0, & z\in(0,z_d^-) \cup (z_d^+,\infty) \\ R=\Delta, & z\in[z_d^-,z_d^+], \end{array} \right.}& \label{eqn:h36} \\
  & E_z = 0 \mbox{ on } z= z_d^+,z_d^-, \quad R\in(0,\Delta), & \label{eqn:h37} \\
  & E\rightarrow 1 \mbox{ as } (R^2 + z^2)\rightarrow\infty \mbox{ uniformly in } \omega, & \label{eqn:h38}
\end{eqnarray}
Here, using the same notation as in~\cite{NL24}, $E=E(R,z)\in[E_s,1]$ is the voidage field (representing the volume of fluid per unit spatial volume), with $0<E_s\ll 1$ being the packing voidage of the particles, and the domain $\omega$ is defined by
\[
  \omega = \{ (R,z): \, z>0, \, R>0 \} \backslash \{ (R,z): \, z_d^- \leq z \leq z_d^+, \, 0 < R\leq \Delta \}.
\]
The effect of the helicopter is modelled as generating an incompressible, inviscid fluid flow field represented by a half-space fluid dipole located at $R=0$, $z=z_d>0$ (in dimensionless variables, with $z_d$ measuring the ratio of the hovering height of the helicopter rotor to the rotor blade radius), and a uniform fluid line vortex aligned along the positive z-axis. The flow takes place in the half-space region $z>0$ and exterior to a small cylindrical shell, where $\Delta>0$ represents the dimensionless length and radius of the cylindrical shell which is aligned with the $z$-axis and with the dipole at its centre, and measures the ratio of the helicopter rotor core radius to the rotor blade radius. In addition,
\[
  z_d^+ = z_d + \frac{\Delta}{2}, \quad z_d^- = z_d - \frac{\Delta}{2},
\]
and the boundary data is given in terms of the continuous and piecewise smooth function $g:[0,\infty)\rightarrow\R$, defined by,
\begin{equation}
  g(R) = \left\{ \begin{array}{ll}
                E_s, & \mbox{for }|\nabla_h \bar{\phi}(R,0)|^2 > (\gamma E_s)^{-1}, \\
                \displaystyle{\gamma^{-1}\left( \frac{9z_d^2 R^2}{4\pi^2(R^2+z_d^2)^5} + \frac{\Omega^2}{4\pi^2 R^2} \right)^{-1}}, & \mbox{for }\gamma^{-1}\leq|\nabla_h \bar{\phi}(R,0)|^2\leq (\gamma E_s)^{-1}, \\
                1, & \mbox{for }|\nabla_h \bar{\phi}(R,0)|^2<\gamma^{-1}. \\
                \end{array} \right.
   \label{eqn:h25}
\end{equation}
Here, again using the notation of~\cite{NL24},
\begin{equation}
  |\nabla_h \bar{\phi}(R,0)|^2 = \frac{9z_d^2 R^2}{4\pi^2(R^2+z_d^2)^5} + \frac{\Omega^2}{4\pi^2 R^2},
  \label{eqn:h20}
\end{equation}

\[
  \bar{a}(R,z) = \frac{9}{2\alpha} a(R,z), \quad \bar{b}(R,z) = \frac{9}{2\alpha} b(R,z),
\]
for $(R,z)\in \omega$, where
\begin{eqnarray}
  a(R,z) & = & \displaystyle{\frac{3R}{4\pi} \left[\frac{(z+z_d)}{(R^2 + (z + z_d)^2)^{5/2}} - \frac{(z-z_d)}{(R^2 + (z - z_d)^2)^{5/2}} \right]}, \label{eqn:h18} \\
  b(R,z) & = & \displaystyle{\frac{3}{4\pi} \left[\frac{(z+z_d)^2}{(R^2 + (z + z_d)^2)^{5/2}} - \frac{(z-z_d)^2}{(R^2 + (z - z_d)^2)^{5/2}} \right]} \nonumber \\
         &   & \quad + \displaystyle{\frac{1}{4\pi} \left[\frac{1}{(R^2 + (z - z_d)^2)^{3/2}} - \frac{1}{(R^2 + (z + z_d)^2)^{3/2}} \right]}. \label{eqn:h19}
\end{eqnarray}
The full derivation of this boundary value problem, from fundamental first principles, is described in~\cite{NL24}, together with a complete discussion of the physical relevance of each of the six dimensionless parameters appearing in this boundary value problem, namely, $E_s$, $\alpha$, $\gamma$, $\Omega$, $z_d$ and $\Delta$.  As discussed in~\cite[\S6]{NL24}, these parameters typically take values as follows:  the parameter $\Delta$ (measuring the ratio of the helicopter rotor core radius to the rotor blade radius) typically satisfies $\Delta\approx 10^{-2}$ (in \S\ref{sec:opacity} and \S\ref{sec:deformation} we take $\Delta=0.04$ throughout); the parameter $z_d$ (measuring the ratio of the hovering height of the helicopter rotor to the rotor blade radius) typically satisfies $z_d\approx10^{-1}$--$10^1$; for the fluid flow the ratio of the hovering helicopter induced swirl velocity to the induced downwash velocity is measured by $\Omega$, and typically we have $\Omega\approx10^{-1}$--$10^0$; the parameter $E_s$ is the particle packing voidage, and for sand in air this is typically of magnitude $10^{-2}$ (and we will take this value throughout); the ratio of the lift force per unit volume to the gravity force per unit volume on particles in the interfacial layer is given by $\gamma$, and we typically have $\gamma\approx10^{2}$--$10^3$; finally, $\alpha$ measures the ratio of the particle collisional pressure to the drag force in the particle flow phase and we tentatively estimate, for sand fluidized in air, that $\alpha\approx10^{-1}$ (and we will take this value throughout). These values are derived using typical helicopter parameters as recorded, for example, in \cite{WaWhKeMcGiDo08,GLG,PoMiPaPeVe20,TaGaBaLiZhHu21,TGZWCW,TaYoHeYuWa22,LSXW,LXJHL}, together with the standard properties of air and sand.  For further details of the modelling, we refer to~\cite{NL24}.

\section{Opacity measure}
\label{sec:opacity}
\begin{figure}
   \centering
   \includegraphics[width=0.49\linewidth]{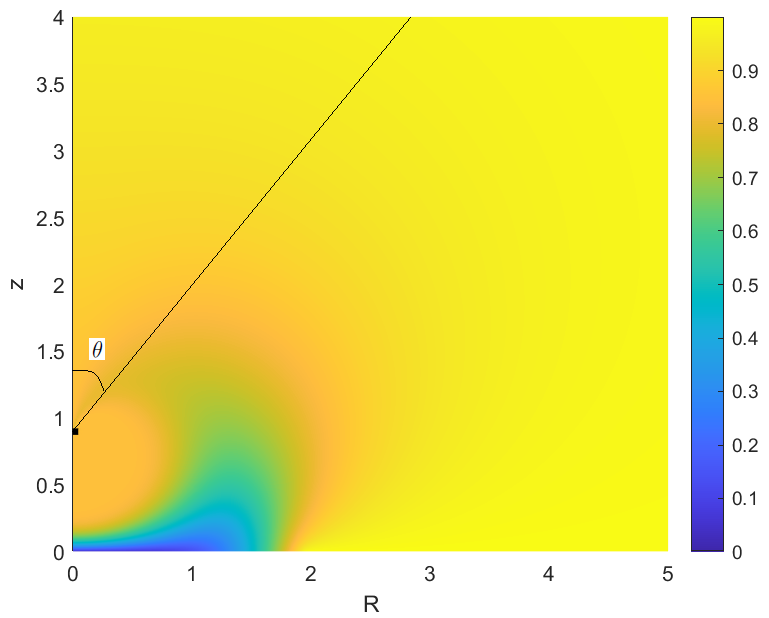}
   \includegraphics[width=0.49\linewidth]{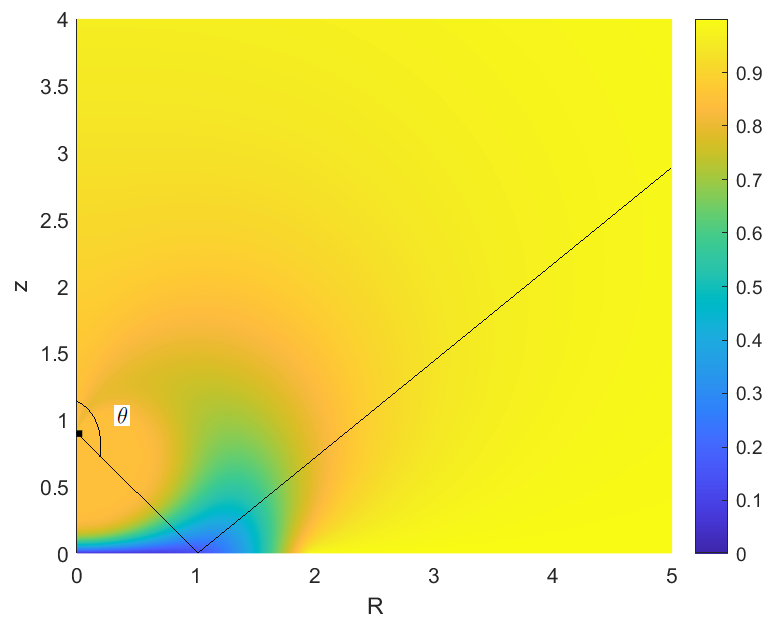}
   \caption{Voidage field $E$, plotted for $\gamma=500$, $\Omega=0.5$, $z_d=0.9$, $R\in[0,5]$, $z\in[0,4]$, showing the line of integration with $\theta$ measured from the downward vertical, for $0\leq\theta\leq\pi/2$ (left) and $\pi/2<\theta\leq\pi$ (right).}
  \label{fig:opacity}
\end{figure}

In relation to the helicopter cloud problem, formulated and analysed in~\cite{NL24}, and reviewed in~\S\ref{sec:model}, we introduce what we will refer to as the \emph{directional opacity measure},
\begin{equation}
  Op(\theta) = \min\left\{\int_{\cal{C}(\theta)} (1-E(R(s),z(s)))\, \rd s, 1\right\} \in [0,1],
  \label{eqn:opacity}
\end{equation}
for each \emph{observation angle} $\theta \in [0,\pi]$. Here $s$ represents arc length from the point at the effective centre of the helicopter, with coordinates $(R,z) = (0,z_d)$, along the integration contour $\cal{C}(\theta)$, which is illustrated in figure~\ref{fig:opacity} (with the voidage field $E$ as shown in the central plot of \cite[figure~10]{NL24}), for the two cases $0\leq\theta\leq\pi/2$ (left plot) and $\pi/2<\theta\leq\pi$ (right plot).  Note that a smaller value of $E$ represents a higher concentration of sand in the air.
This provides a reasonable and normalised measure of the light transmission defect from the far field to the helicopter cab, in each direction $\theta$, due to the presence of the particle cloud. In the context of this measure, $Op(\theta)=0$ implies perfect visibility, corresponding to $E=1$ everywhere along the contour, whilst $Op(\theta)=1$ implies very poor visibility (high opacity and total brownout), corresponding to $E_s<E\ll 1$ on significant parts of the contour. In terms of this measure, we can introduce what we will refer to as the \emph{brownout set}, with definition,
\[ \mathcal{B}  = \left\{\lambda \in [0,1]: Op(\lambda \pi) = 1\right\} \]
which consists of those angles where total brownout is recorded.

To determine $Op(\theta)$ for a given set of model parameters, we first calculate the voidage field $E$ from the boundary value problem (\ref{eqn:h34})--(\ref{eqn:h38}) using the numerical method described in \cite[Appendix~B]{NL24}. For each chosen $\theta \in [0,\pi]$, we then evaluate the line integral in (\ref{eqn:opacity}) using standard quadrature, taking care to ensure we have taken enough quadrature points to guarantee several decimal places of accuracy.

\begin{figure}
   \centering
   \includegraphics[width=0.458\linewidth]{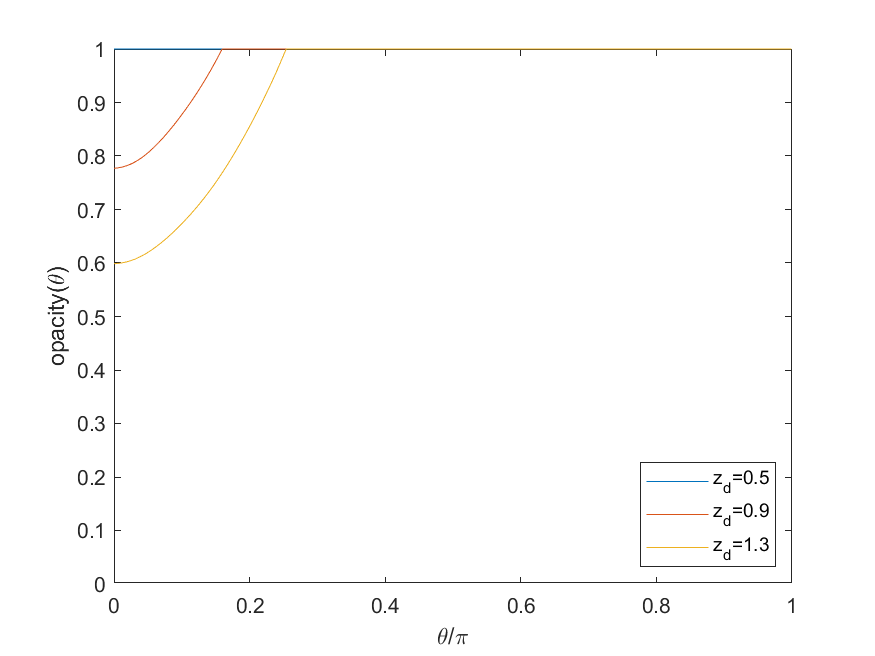}
   \includegraphics[width=0.458\linewidth]{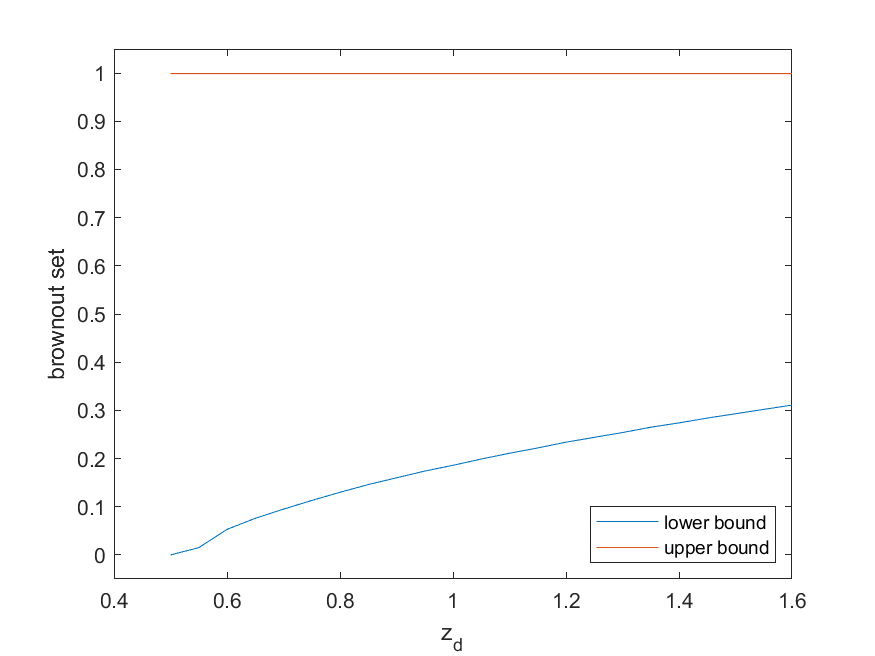}
   \caption{Directional Opacity Measure, plotted for $\gamma=1000$, $\Omega=1.0$, $z_d=0.5, 0.9, 1.3$ (left), and lower and upper bounds for $\mathcal{B}$, plotted against $z_d\in[0.5,1.6]$ (right).  Here, $\mathcal{B}  = [0,1]$, for $z_d=0.5$, $\mathcal{B}  = [0.16,1]$, for $z_d=0.9$, and $\mathcal{B}  = [0.25,1]$, for $z_d=1.3$.}
  \label{fig:opacity_results2}
\end{figure}

\begin{figure}
   \centering
   \includegraphics[width=0.458\linewidth]{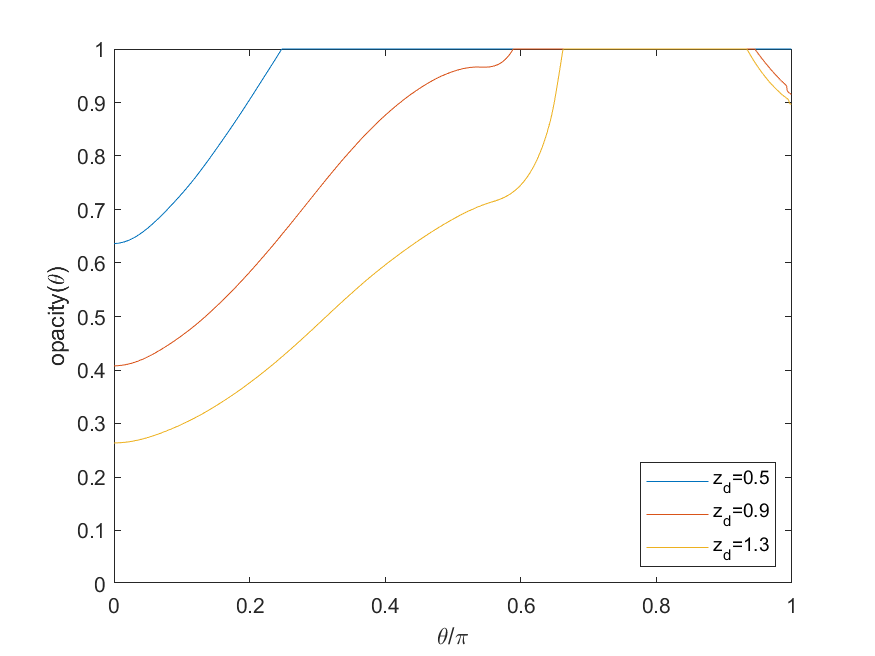}
   \includegraphics[width=0.458\linewidth]{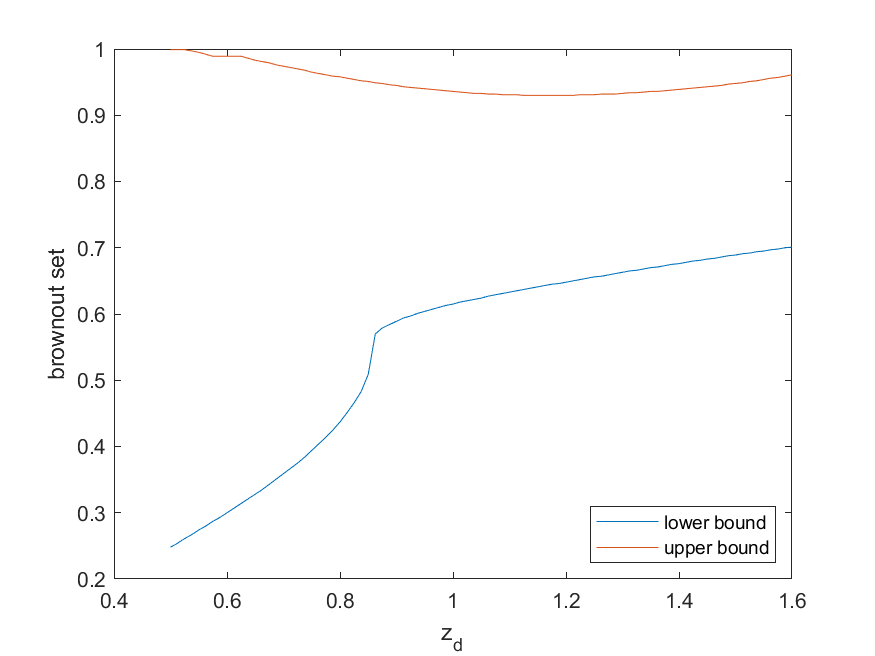}
   \caption{Directional Opacity Measure, plotted for $\gamma=1000$, $\Omega=0.5$, $z_d=0.5, 0.9, 1.3$ (left), and lower and upper bounds for $\mathcal{B}$, plotted against $z_d\in[0.5,1.6]$ (right).  Here, $\mathcal{B}  = [0.25,1.00]$, for $z_d=0.5$, $\mathcal{B}  = [0.59,0.95]$, for $z_d=0.9$, and $\mathcal{B}  = [0.66,0.93]$, for $z_d=1.3$.}
  \label{fig:opacity_results3}
\end{figure}

\begin{figure}
   \centering
   \includegraphics[width=0.458\linewidth]{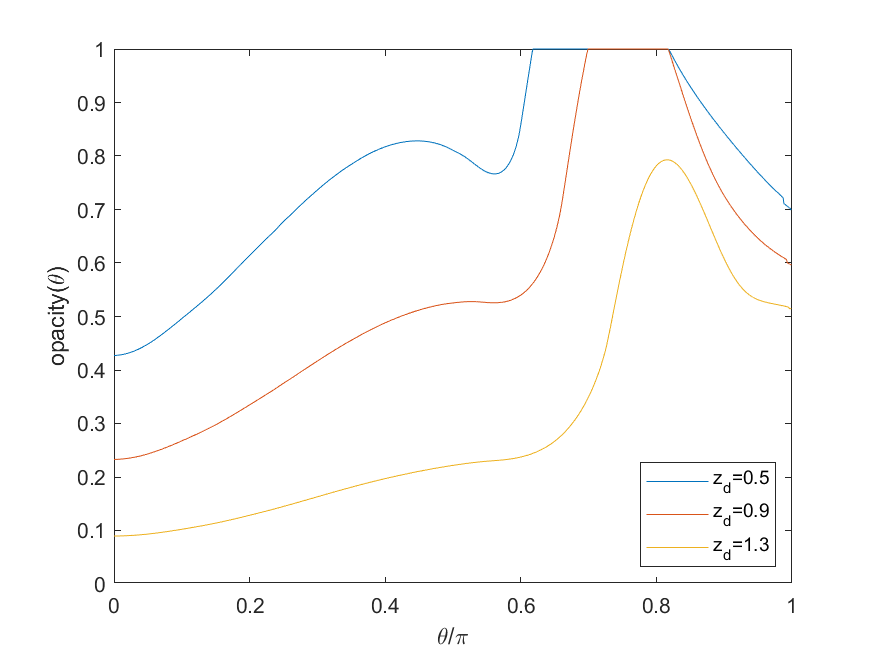}
   \includegraphics[width=0.458\linewidth]{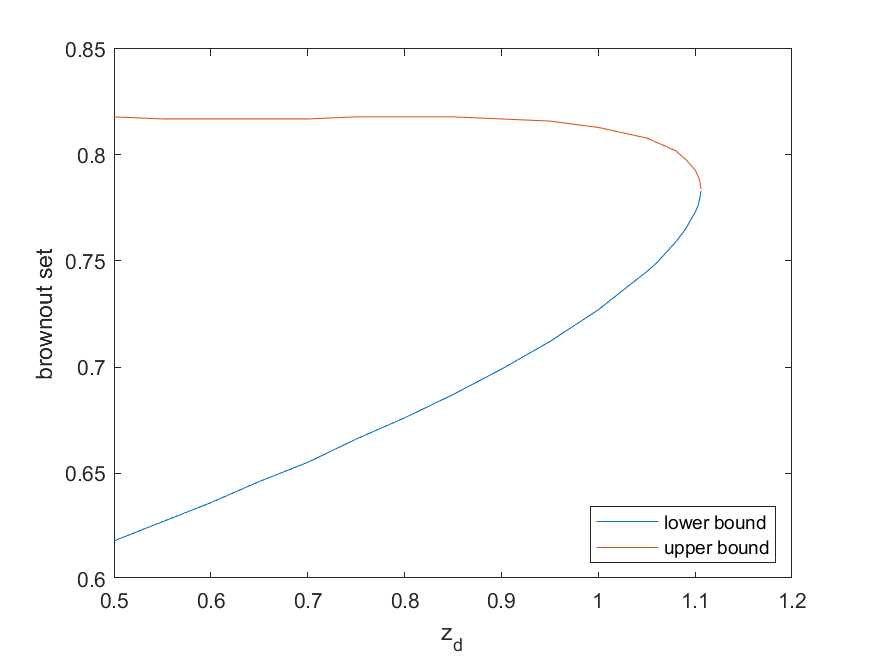}
   \caption{Directional Opacity Measure, plotted for $\gamma=1000$, $\Omega=0.1$, $z_d=0.5, 0.9, 1.3$ (left), and lower and upper bounds for $\mathcal{B}$, plotted against $z_d\in[0.5,1.2]$ (right).  Here, $\mathcal{B}  = [0.62,0.82]$, for $z_d=0.5$, $\mathcal{B}  = [0.70,0.82]$, for $z_d=0.9$, and $\mathcal{B}$ empty, for $z_d=1.3$.}
  \label{fig:opacity_results4}
\end{figure}

\begin{figure}
   \centering
   \includegraphics[width=0.458\linewidth]{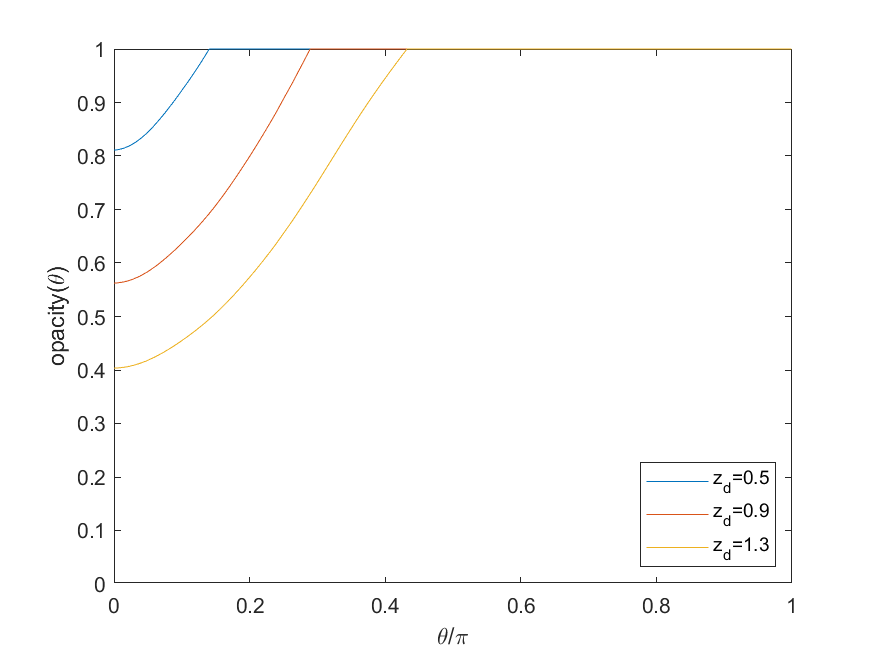}
   \includegraphics[width=0.458\linewidth]{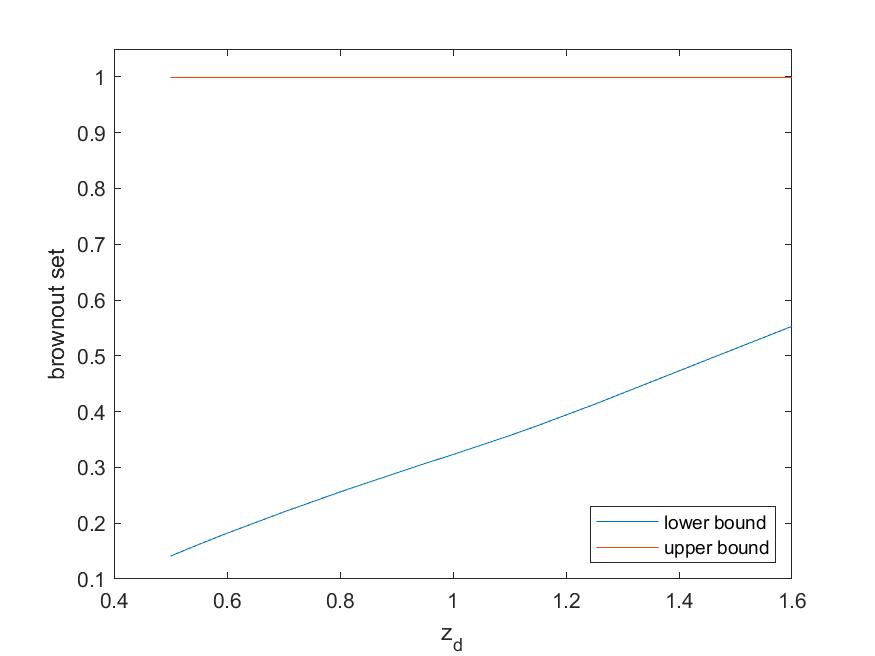}
   \caption{Directional Opacity Measure, plotted for $\gamma=500$, $\Omega=1.0$, $z_d=0.5, 0.9, 1.3$ (left), and lower and upper bounds for $\mathcal{B}$, plotted against $z_d\in[0.5,1.6]$ (right).  Here, $\mathcal{B}  = [0.14,1]$, for $z_d=0.5$, $\mathcal{B}  = [0.29,1]$, for $z_d=0.9$, and $\mathcal{B}  = [0.43,1]$, for $z_d=1.3$.}
  \label{fig:opacity_results5}
\end{figure}

\begin{figure}
   \centering
   \includegraphics[width=0.458\linewidth]{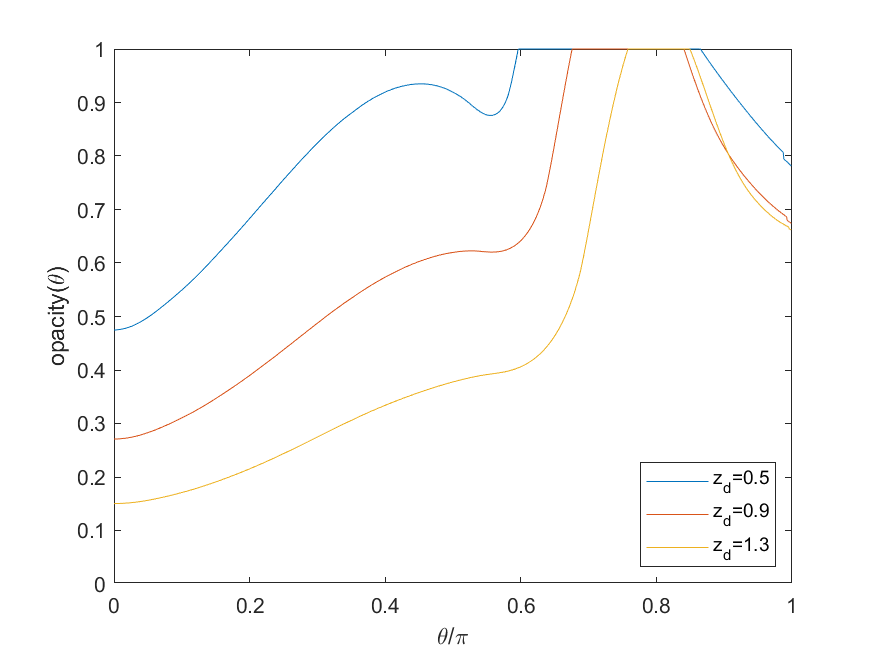}
   \includegraphics[width=0.458\linewidth]{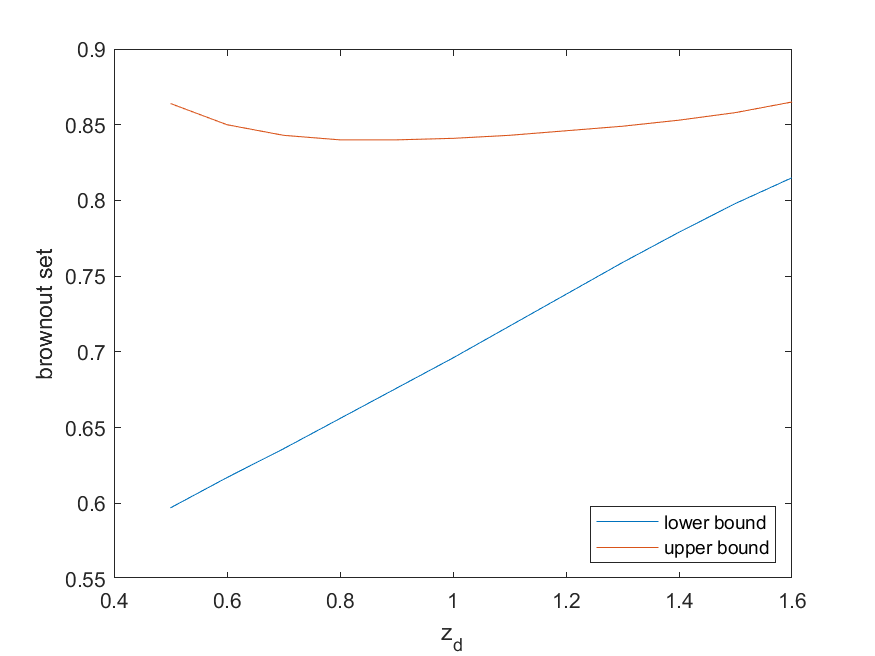}
   \caption{Directional Opacity Measure, plotted for $\gamma=500$, $\Omega=0.5$, $z_d=0.5, 0.9, 1.3$ (left), and lower and upper bounds for $\mathcal{B}$, plotted against $z_d\in[0.5,1.6]$ (right). Here, $\mathcal{B}  = [0.60,0.86]$, for $z_d=0.5$, $\mathcal{B}  = [0.68,0.84]$, for $z_d=0.9$, and $\mathcal{B}  = [0.76,0.85]$, for $z_d=1.3$, with the slightly counter-intuitive implication that there is a small range of radial directions where visibility is better when the helicopter is hovering at a lower height.}
  \label{fig:opacity_results6}
\end{figure}

\begin{figure}
   \centering
   \includegraphics[width=0.458\linewidth]{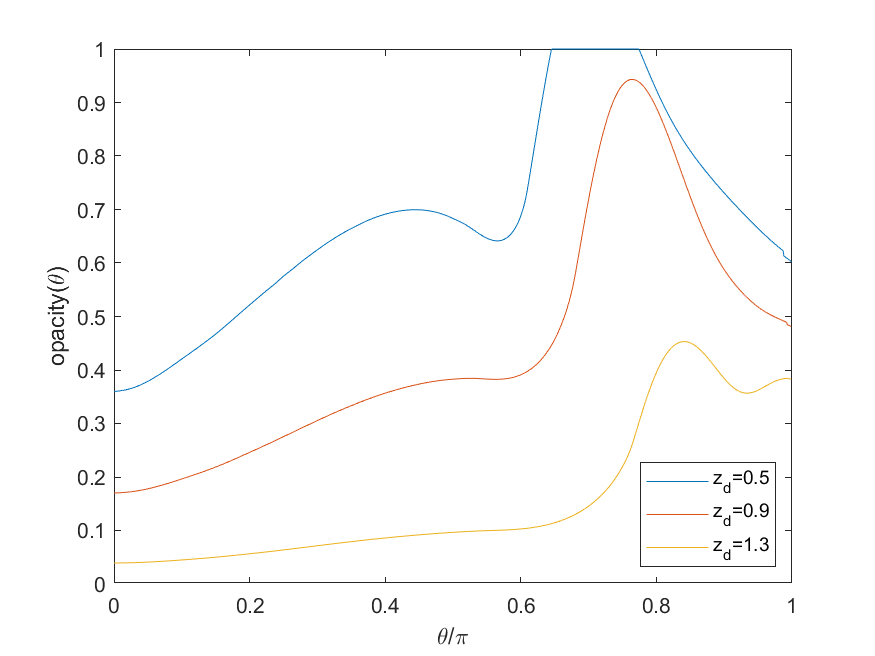}
   \includegraphics[width=0.458\linewidth]{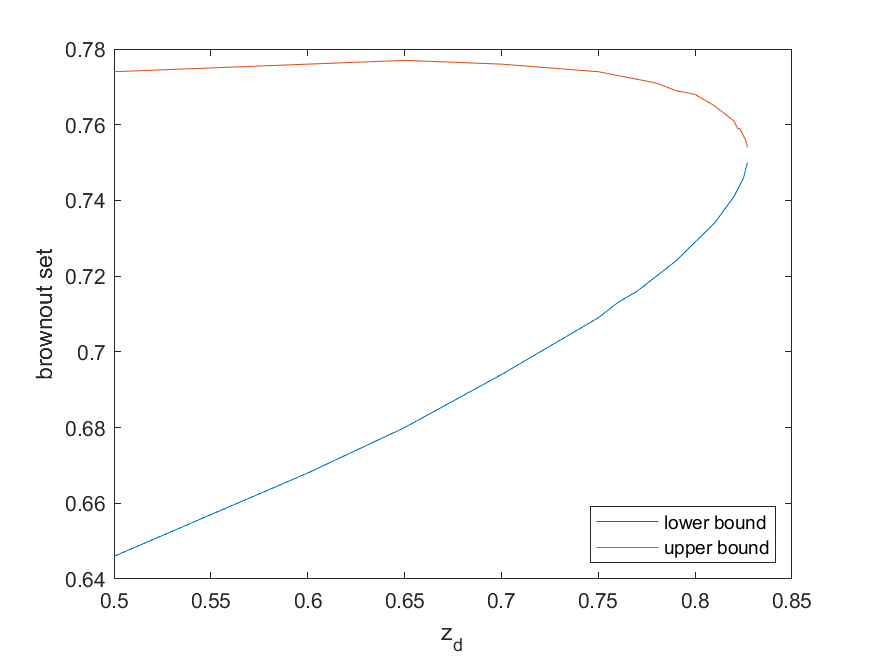}
   \caption{Directional Opacity Measure, plotted for $\gamma=500$, $\Omega=0.1$, $z_d=0.5, 0.9, 1.3$ (left), and lower and upper bounds for $\mathcal{B}$, plotted against $z_d\in[0.5,0.85]$ (right).  Here, $\mathcal{B}  = [0.65,0.77]$ for $z_d=0.5$, and $\mathcal{B}$ empty for both $z_d=0.9$ and $z_d=1.3$.}
  \label{fig:opacity_results7}
\end{figure}

\begin{figure}
   \centering
   \includegraphics[width=0.458\linewidth]{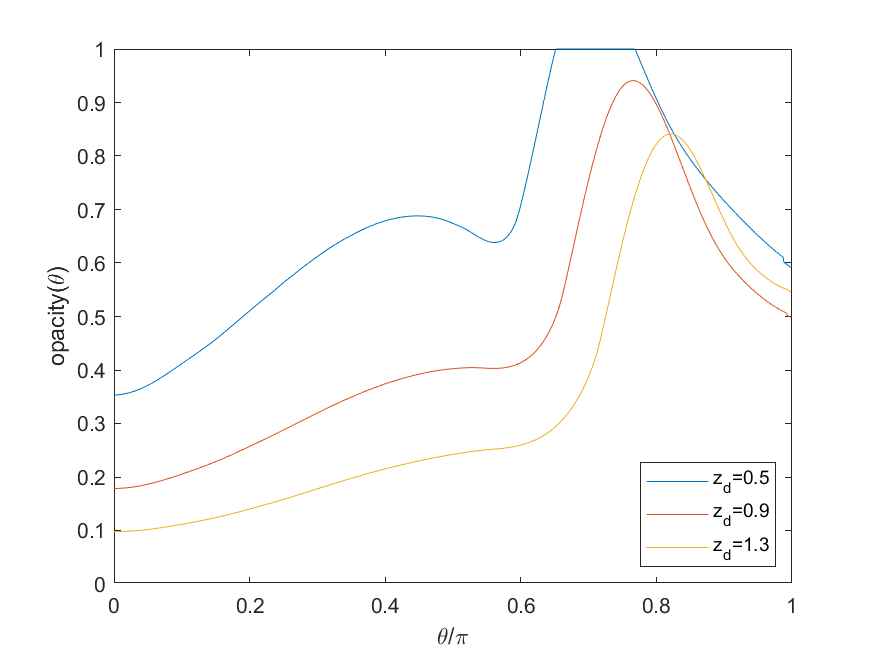}
   \includegraphics[width=0.458\linewidth]{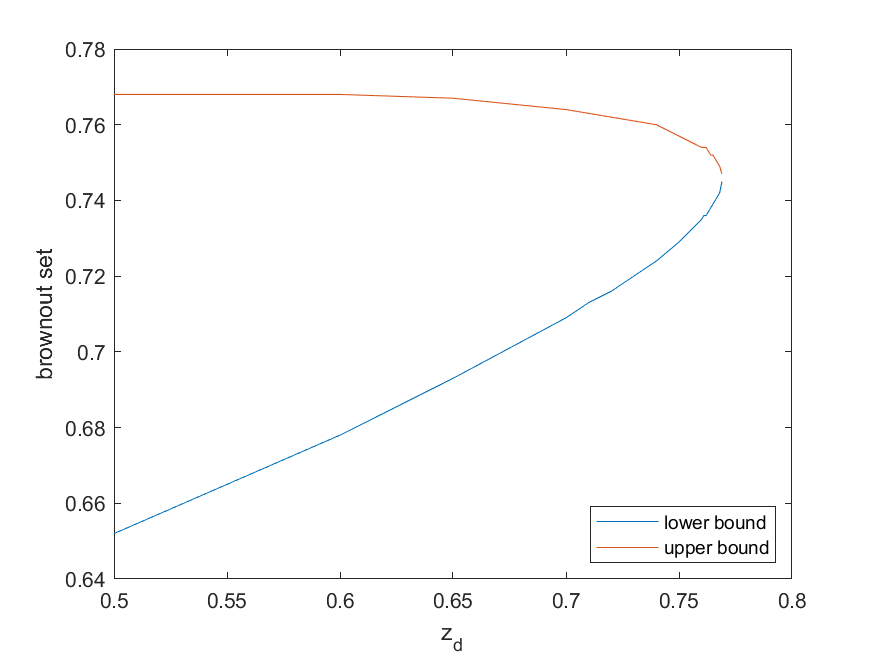}
   \caption{Directional Opacity Measure, plotted for $\gamma=100$, $\Omega=1.0$, $z_d=0.5, 0.9, 1.3$ (left), and lower and upper bounds for $\mathcal{B}$, plotted against $z_d\in[0.5,0.8]$ (right).  Here, $\mathcal{B}  = [0.65,0.77]$ for $z_d=0.5$, and $\mathcal{B}$ is empty for both $z_d=0.9$ and $z_d=1.3$.}
  \label{fig:opacity_results8}
\end{figure}

\begin{figure}
   \centering
   \includegraphics[width=0.458\linewidth]{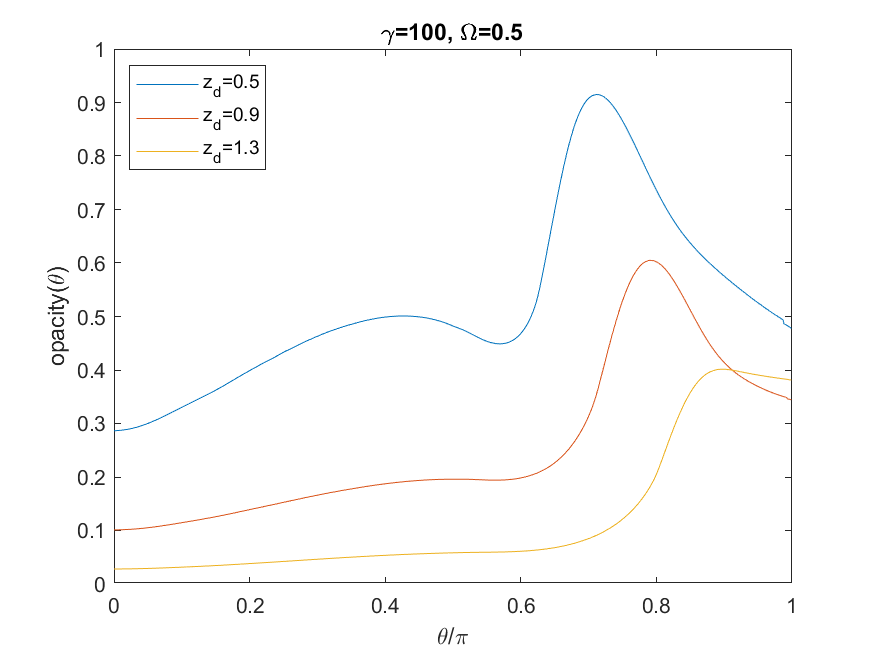}
   \includegraphics[width=0.458\linewidth]{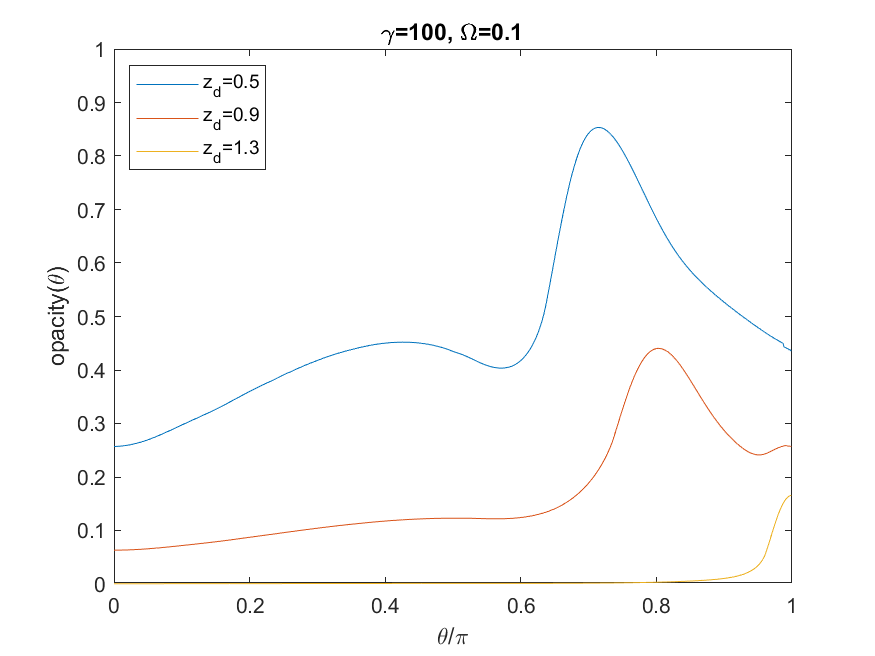}
   \caption{Directional Opacity Measure, plotted for $\gamma=100$, $z_d=0.5, 0.9, 1.3$, and $\Omega=0.5$ (left) and $\Omega=0.1$ (right).  Here the brownout set $\mathcal{B}$ is empty for each value of $z_d$.}
  \label{fig:opacity_results9and10}
\end{figure}

For the example shown in figure~\ref{fig:opacity}, we might expect that $Op(\theta)$ will be small when $\theta\ll\pi/2$, with the value then increasing as $\theta$ grows, up to a maximum as the line of integration passes through the blue part of the figure (where $E_s<E\ll1$), before decreasing again as $\theta$ becomes closer to $\pi$.  This is seen on the red line on the left of figure~\ref{fig:opacity_results6}, which corresponds to the voidage field shown in figure~\ref{fig:opacity}, and from which we can estimate the corresponding brownout set as $\mathcal{B} = [0.68,0.84]$.  Opacity results, for each combination of $\gamma=1000, 500, 100$, $\Omega=1,0.5,0.1$ and $z_d=1.3, 0.9, 0.5$, (corresponding to the plots of voidage field $E$ in \cite[figures~9--11]{NL24}), each of which is qualitatively (though not quantitatively) comparable to that shown in figure~\ref{fig:opacity} are shown in figures~\ref{fig:opacity_results2}--\ref{fig:opacity_results9and10}, with the corresponding brownout set for each value of $z_d$ recorded in the respective caption.  We also show the upper and lower bounds of the brownout set for $z_d\in[0.5,1.6]$ on the right of figures~\ref{fig:opacity_results2}--\ref{fig:opacity_results8} (the cases where the brownout set is not empty).

For each plot, we see the expected behaviour when considered in light of the results in \cite[figures~9--11]{NL24}.  As explained in \cite[\S7]{NL24}, we generally see more sand particles being entrained into the fluidized region (i.e., increased opacity and lower visibility) when $\gamma$ and $\Omega$ are larger, since large $\gamma$ means that lift dominates gravity in the transition layer, and larger $\Omega$ means a higher ratio of swirl velocity to downdraft velocity (recall the definition of $g$, (\ref{eqn:h25})).

When both $\gamma$ and $\Omega$ are large, in particular as seen in figures~\ref{fig:opacity_results2}, \ref{fig:opacity_results3} and~\ref{fig:opacity_results5}, we have high opacity (low visibility) for all $\theta$ sufficiently large, with zero visibility for a larger range of values of $\theta$ the lower the value of $z_d$ (i.e., the closer the helicopter is to the ground).

For figures~\ref{fig:opacity_results4}, \ref{fig:opacity_results6} (corresponding to the voidage field seen in figure~\ref{fig:opacity}), \ref{fig:opacity_results7} and~\ref{fig:opacity_results8}, we see a broadly similar pattern in each case, with $ Op(\theta)$ small when $0 \le \theta\ll\pi/2$, then generally (though not uniformly) increasing as $\theta$ grows, with a region of zero visibility (maximum opacity) when $z_d$ is sufficiently small, and $Op(\theta)$ then decreasing again as $\theta$ increases further towards $\pi$.  In each case, opacity is generally greater when $z_d$ is smaller, though in figures~\ref{fig:opacity_results6} and~\ref{fig:opacity_results8} (and also on the left of figure~~\ref{fig:opacity_results9and10}) there is a small range of values of $\theta$ where $Op(\theta)$ is greater for $z_d=1.3$ than it is for $z_d=0.9$.  In each case, particularly noticeable when $z_d$ is smaller, we see a small drop in opacity when $\theta\approx\pi$;  this is because as $\theta$ approaches $\pi$, the line of integration passes back through the cylindrical shell with height and radius $\Delta$ and with $z_d$ at its centre (the small black box that can be seen in figure~\ref{fig:opacity}), with no contribution to the integral in~(\ref{eqn:opacity}) from the section of the line within the shell.

In figure~\ref{fig:opacity_results9and10} (corresponding to smaller values of $\gamma$ and $\Omega$) we see qualitatively comparable behaviour to that described above, but with no region of maximum opacity (zero visibility) - in these cases, and noting the discussion above, there are fewer sand particles entrained into the fluidized region than in the other examples (as seen in \cite[figure~11]{NL24}).

A final interesting, and somewhat unanticipated, feature of the opacity graphs is the non-monotonic nature of a significant number of the curves within a range of observation angles of practical interest for pilot vision purposes, say $\theta \in [\pi/3,2\pi/3]$. It might have been expected that the curves would simply have increasing opacity  with increasing observation angle. However, at a given value of $\gamma$, when $\Omega$ becomes sufficiently small (but without being extreme) then for $z_d = 0.5$ or below, we see a weak maximum, followed by a stronger minimum, and thereafter an expected increase followed by a second stronger maximum and a final decrease, with the two maxima and the intervening minimum all in the range  of observation angles of practical interest. We will refer to the local maxima as representing  \emph{corridors of lower vision} (CLV) and the corresponding deeper local minimum as representing \emph{a corridor of higher vision} (CHV). The strength of both the (CLV) and (CHV) increase with decreasing $\Omega$ and/or $\gamma$, and decreasing $z_d$. Referring back to the voidage field contour plots, an example of which is shown in figure~\ref{fig:opacity}, this phenomenon can be seen to be due to the subtle nonlinear balance between the low voidage values in the curved `horn-like' region, and the associated local thickness of this region, as it is traversed by the opacity contour $\mathcal{C}(\theta)$. Small, but nontrivial, gain or loss of vision is achieved along these respective corridors.

We now consider the lower and upper bounds of the brownout set.  In most cases considered (figures~\ref{fig:opacity_results2}, \ref{fig:opacity_results4}, \ref{fig:opacity_results5},  \ref{fig:opacity_results8} in particular), the lower bound increases and the upper bound either remains approximately constant or decreases, as $z_d$ increases, as we might expect (i.e., better visibility when $z_d$ is greater).  Note that in a number of cases the brownout set disappears:  for $z_d\in[1.10,1.11]$ in figure~\ref{fig:opacity_results4}, for $z_d\in[0.82,0.83]$ in figure~\ref{fig:opacity_results7}, and for $z_d\in[0.76,0.77]$ in figure~\ref{fig:opacity_results8}.  In figure~\ref{fig:opacity_results7}, the upper bound increases very marginally as $z_d$ increases, before decreasing.  However, in figures~\ref{fig:opacity_results3}  and~\ref{fig:opacity_results6} we see somewhat different behaviour.

\begin{figure}
   \centering
   \includegraphics[width=0.458\linewidth]{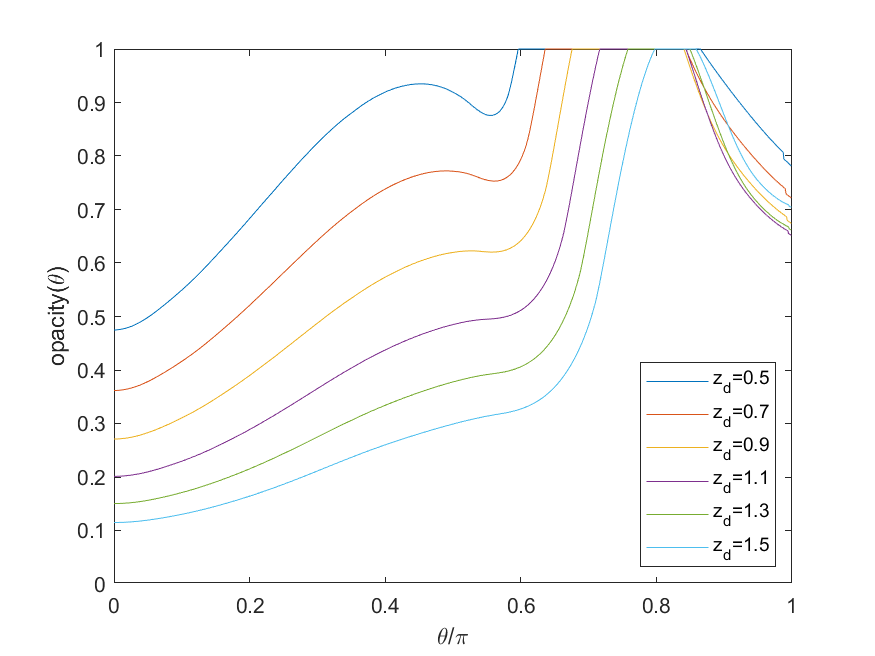}
   \includegraphics[width=0.458\linewidth]{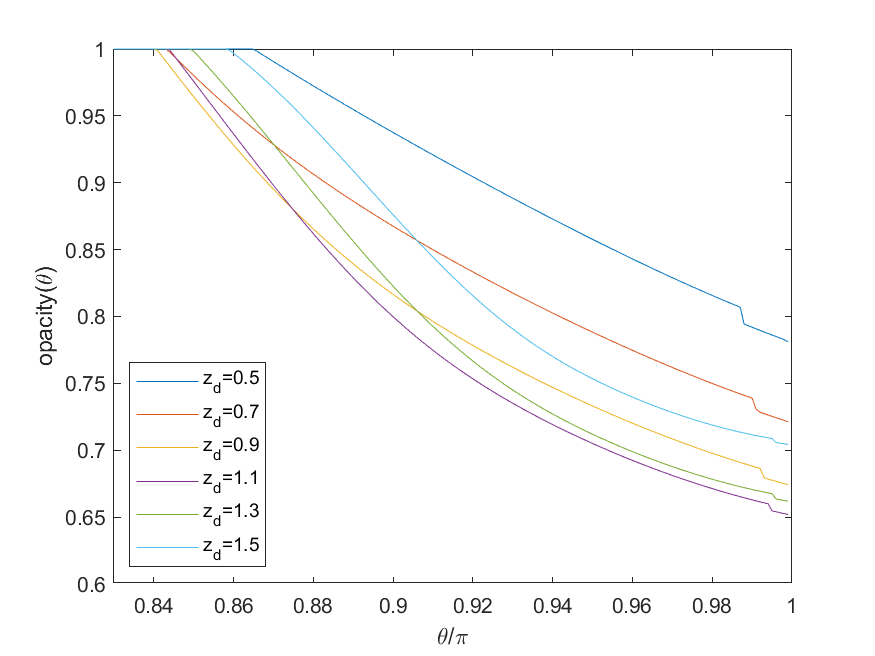}
   \caption{Directional Opacity Measure, plotted for $\gamma=500$, $\Omega=0.5$, $z_d\in[0.5,1.5]$.}
  \label{fig:opacity_results6_critical_zd}
\end{figure}

In figure~\ref{fig:opacity_results6}, we see that the upper bound initially decreases as $z_d$ increases, before then growing again for $z_d$ greater than $z_d\approx0.8$.  To explore this further, we plot the directional opacity measure in this case ($\gamma=500$, $\Omega=0.5$) for a range of values of $z_d$ in figure~\ref{fig:opacity_results6_critical_zd} (compare to figure~\ref{fig:opacity_results6}, where the directional opacity measure is plotted for $z_d=0.5, 0.9, 1.3$).  On the right of figure~\ref{fig:opacity_results6}, we show the same plots but zoomed in on the range $\theta/\pi>0.84$.  It is clear that there are a range of values of $\theta$ for which visibility can deteriorate as $z_d$ increases; in particular, within this range, we see visibility improve as $z_d$ increases from $z_d=0.5$ to $z_d\approx1.1$, before deteriorating as $z_d$ increases to $z_d=1.3$ and then $z_d=1.5$.

\begin{figure}
   \centering
   \includegraphics[width=0.7\linewidth]{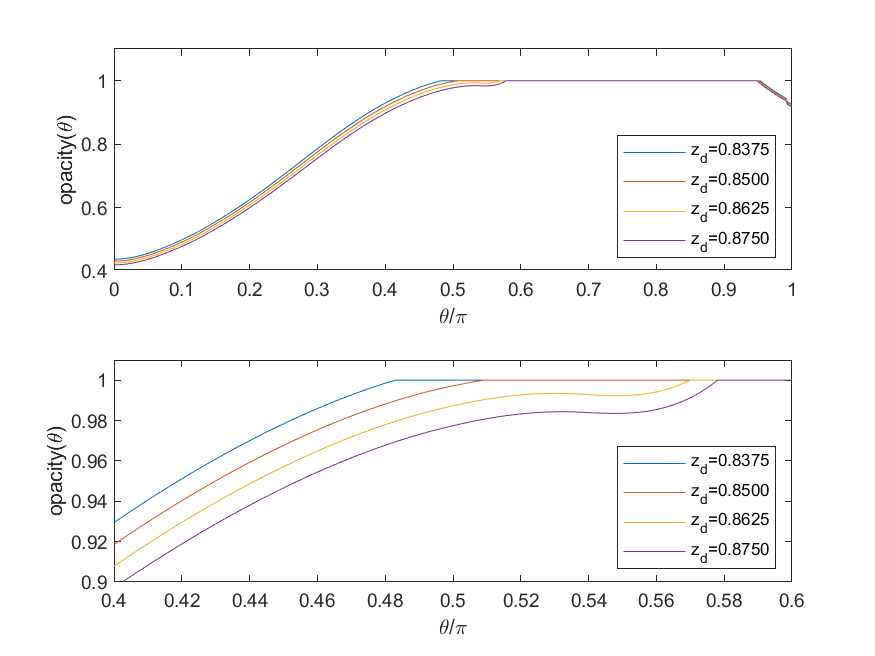}
   \caption{Directional Opacity Measure, plotted for $\gamma=1000$, $\Omega=0.5$, $z_d=0.8375, 0.85, 0.8625, 0.875$.  The lower plot shows the same results as the upper plot, but zoomed in around the critical region in order to show the key solution behaviour more clearly.}
  \label{fig:opacity_results3_critical_zd}
\end{figure}

Finally, in figure~\ref{fig:opacity_results3}, we see a rather unusual pattern of behaviour, for the lower bound in particular (with the upper bound behaving in a manner broadly comparable to that seen in figure~\ref{fig:opacity_results6}).  Here, the lower bound increases very sharply around $z_d\approx0.85$, before flattening off.  To explore this further, we plot the directional opacity measure in this case ($\gamma=1000$, $\Omega=0.5$) over a critical range of values of $z_d$ in figure~\ref{fig:opacity_results3_critical_zd} (compare to figure~\ref{fig:opacity_results3}), with the changing nature of the directional opacity measure as $z_d$ increases from $z_d=0.85$ to $z_d=0.8625$ clearly visible.

We now move on to consider the deformation of the upper surface of the static particle bed due to the entrainment / detrainment of particles into / from the fluidized region above due to the fluid flow induced by the helicopter model.

\section{Surface deformation of the static particle bed}
\label{sec:deformation}
In this section we consider the deformation of the upper surface of the static particle bed again for the model helicopter cloud problem introduced in~\cite{NL24}, and as summarised here in~\S\ref{sec:model}. In particular, for a helicopter hovering steadily over, say, the dimensionless time interval $t\in [0,T]$, the surface displacement can be obtained directly from~\cite{NL24} (see the last equation in Section 5) as the surface deformation formula
\[ \xi(R,t) =  \chi(R) t, \quad t\in [0,T], \]
where, broadly following the notation of~\cite{NL24},
\[ \chi(R) = \left.-\frac{\alpha}{(2-E_s-E)}\right|_{z=0} \times \left. \frac{1}{\bar{\beta}(E)}\right|_{z=0} \times \left. E_z \right|_{z=0}, \quad R\geq 0. \]
We observe that, up to the approximations made in obtaining the model, when the helicopter is hovering steadily, the surface deformation is self similar, with time $t$ simply playing the role of an increasing scale multiplier. Having solved  the boundary value problem (\ref{eqn:h34})--(\ref{eqn:h38}) for the voidage field $E$ using the numerical method described in \cite[Appendix~B]{NL24}, we can then approximate $\left. E_z \right|_{z=0}$ (the derivative on the boundary) using a simple difference formula;  here, we have used the three-point rule (accurate to $O(h^2)$)
\[ E_z(R,0) \approx \frac{-E(R,2h) + 4E(R,h) - 3E(R,0)}{2h}, \]
where $h$ is the mesh width.  Noting that $E_s \le E\le1$ and $0<E_s<1$, and that (see~\cite{NL24})
\[
  \bar{\beta}(E) = \left(\frac{9}{2}\right)\beta(E),
\]
where $\beta(E)\geq 1$ is a decreasing function of $E\in[E_s,1]$, with $\beta(1)=1$, representing the effect of neighbouring particles on the Stokes drag (see \cite[(3.10)]{NL24}), it is clear that it will be of particular interest to see where $E_z$ changes sign.  Following \cite[(6.28)]{NL24} we have, as a first approximation, the simple linear form
\[
  \bar{\beta}(E) = \bar{\beta}_0 - (\bar{\beta}_0 - 9/2) E,
\]
for $E\in[E_s,1]$, with $\bar{\beta}_0 > 9/2$ a material constant.  As explained in~\cite{NL24} we anticipate that, in general, the variation in $\bar{\beta}(E)$ over $E\in[E_s,1]$ will be small, so we write
\[
  \bar{\beta}_0 = \frac{9}{2} + \delta \beta_0,
\]
with $0<\delta\beta_0\ll 1$, and hence,
\[ \bar{\beta}(E) = \frac{9}{2} + \delta \beta_0 (1-E). \]
In all results below we have taken $\delta\beta_0=0.01$.

\begin{figure}
   \centering
   \includegraphics[width=0.49\linewidth]{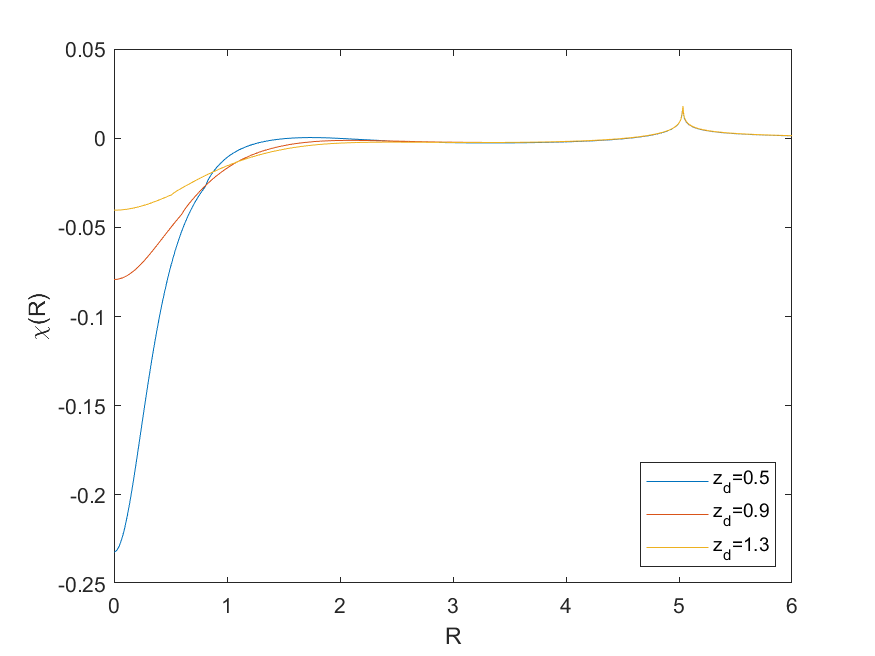}
   \includegraphics[width=0.49\linewidth]{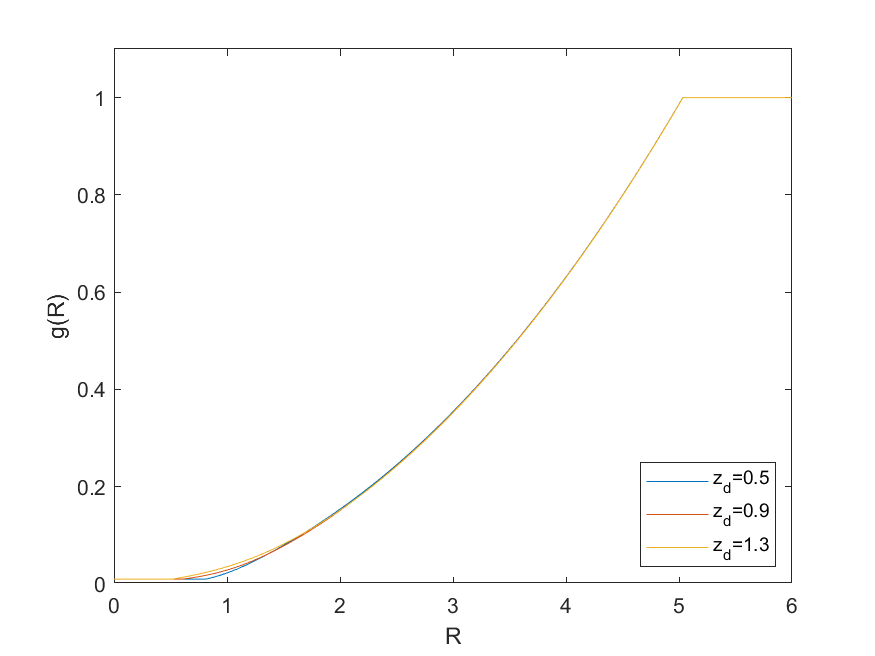}
   \caption{Surface deformation (left), and boundary data (right), plotted for $\gamma=1000$, $\Omega=1.0$, $z_d=1.3, 0.9, 0.5$.}
  \label{fig:sd_results2}
\end{figure}

\begin{figure}
   \centering
   \includegraphics[width=0.49\linewidth]{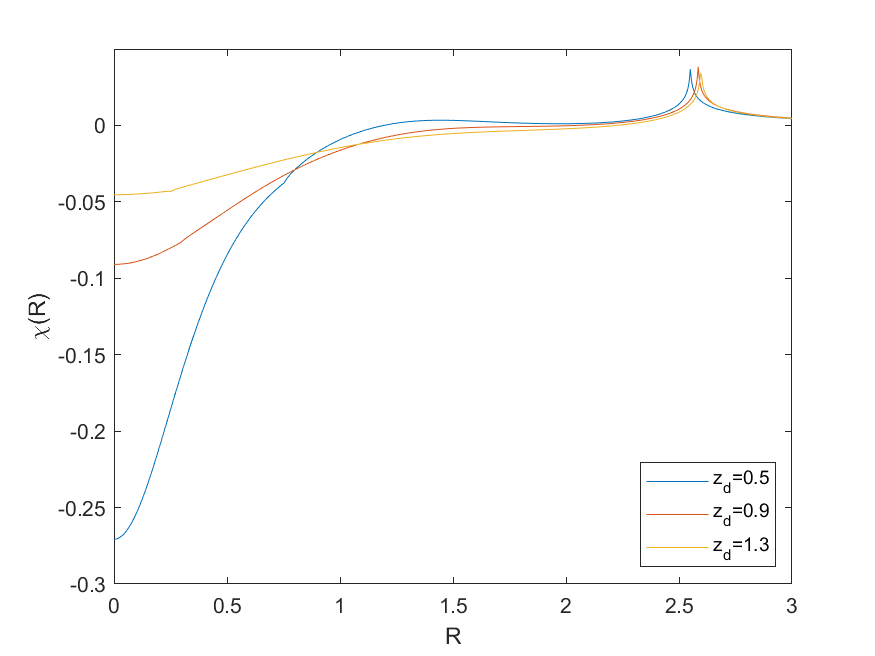}
   \includegraphics[width=0.49\linewidth]{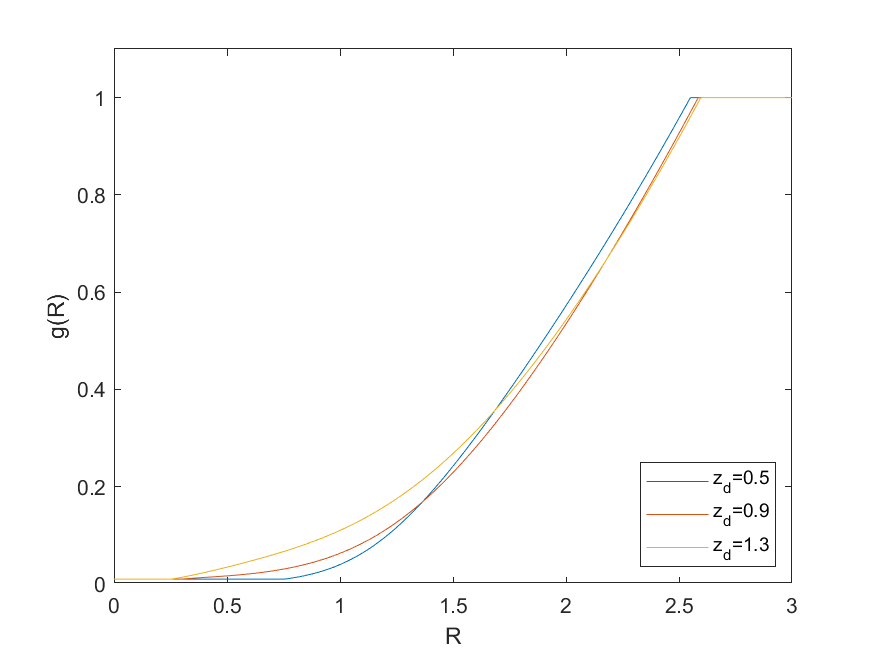}
   \caption{Surface deformation (left), and boundary data (right), plotted for $\gamma=1000$, $\Omega=0.5$, $z_d=1.3, 0.9, 0.5$.}
  \label{fig:sd_results3}
\end{figure}

\begin{figure}
   \centering
   \includegraphics[width=0.49\linewidth]{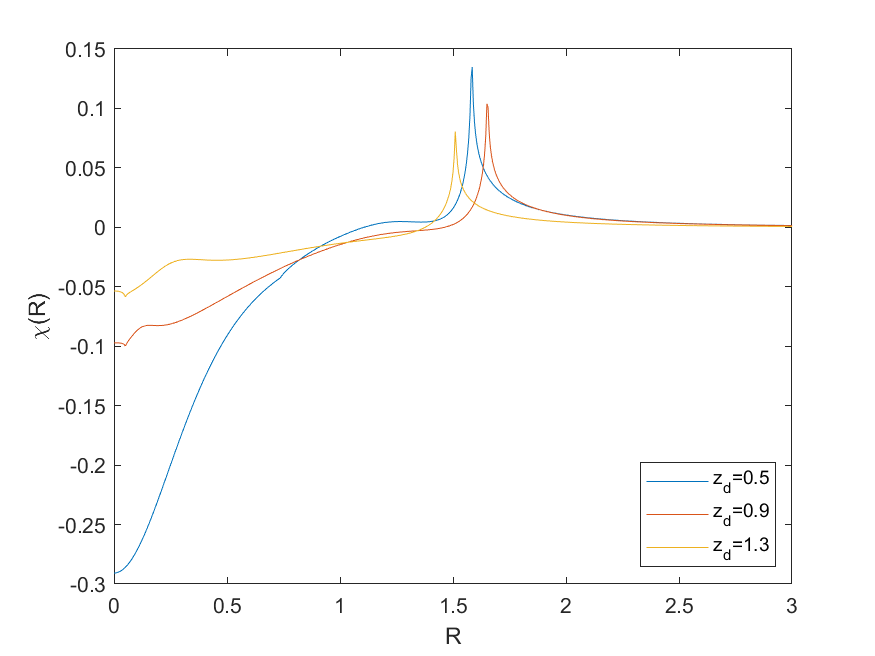}
   \includegraphics[width=0.49\linewidth]{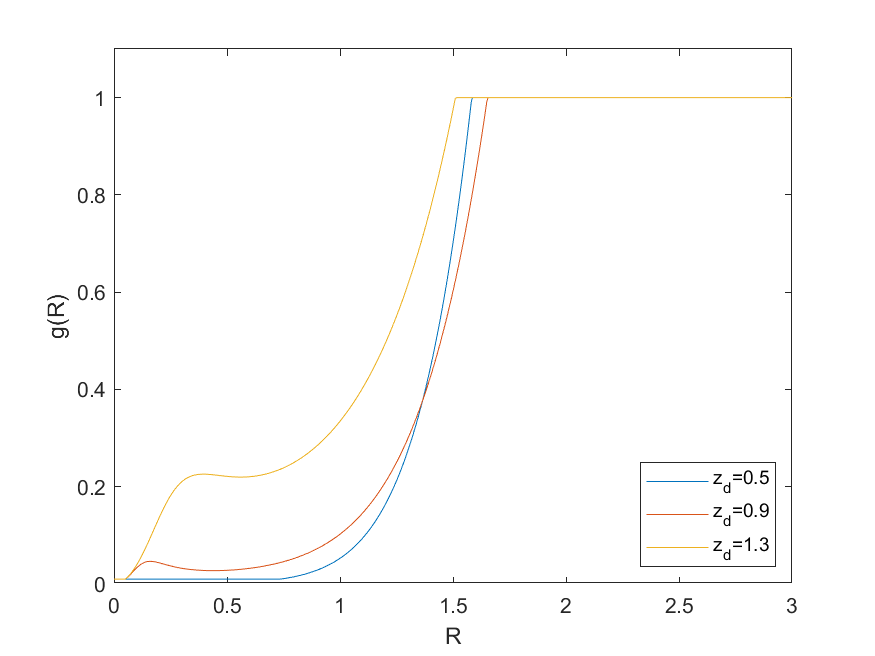}
   \caption{Surface deformation (left), and boundary data (right), plotted for $\gamma=1000$, $\Omega=0.1$, $z_d=1.3, 0.9, 0.5$.}
  \label{fig:sd_results4}
\end{figure}

\begin{figure}
   \centering
   \includegraphics[width=0.49\linewidth]{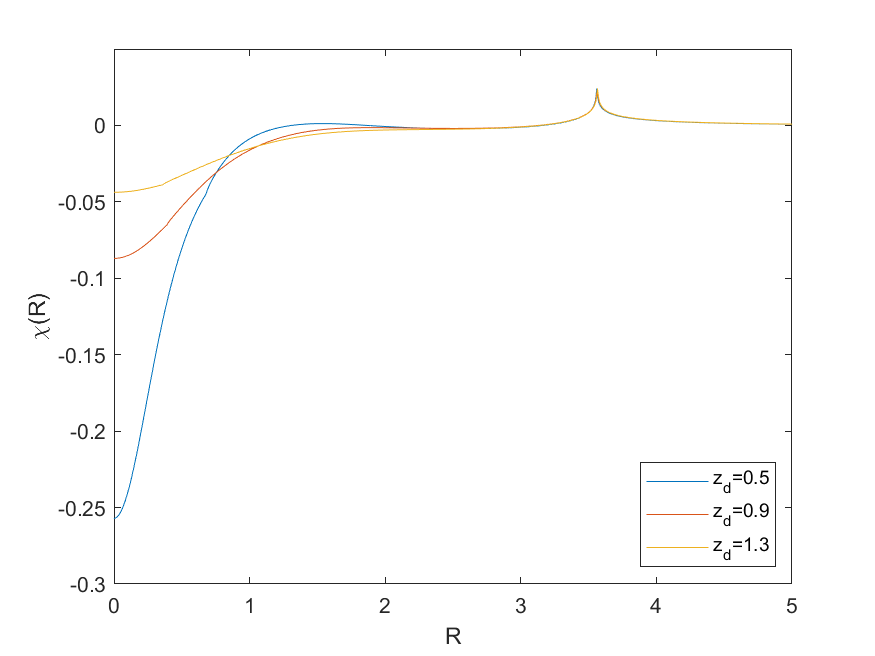}
   \includegraphics[width=0.49\linewidth]{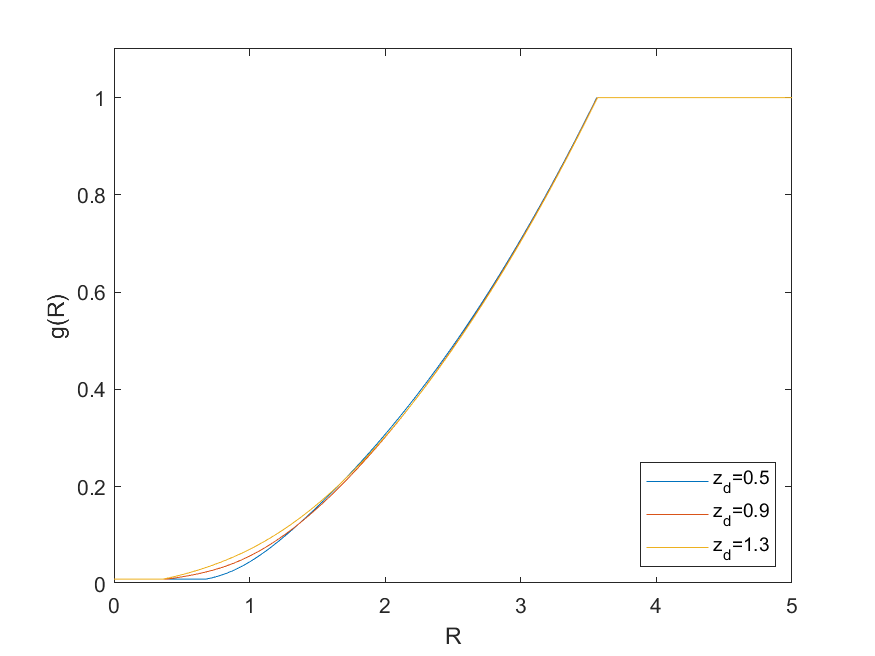}
   \caption{Surface deformation (left), and boundary data (right), plotted for $\gamma=500$, $\Omega=1.0$, $z_d=1.3, 0.9, 0.5$.}
  \label{fig:sd_results5}
\end{figure}

\begin{figure}
   \centering
   \includegraphics[width=0.49\linewidth]{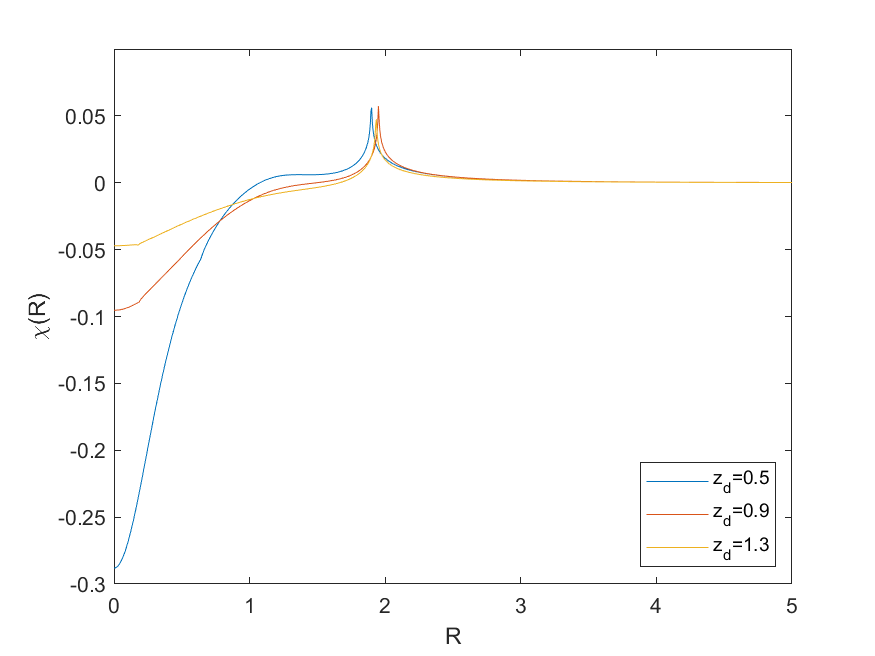}
   \includegraphics[width=0.49\linewidth]{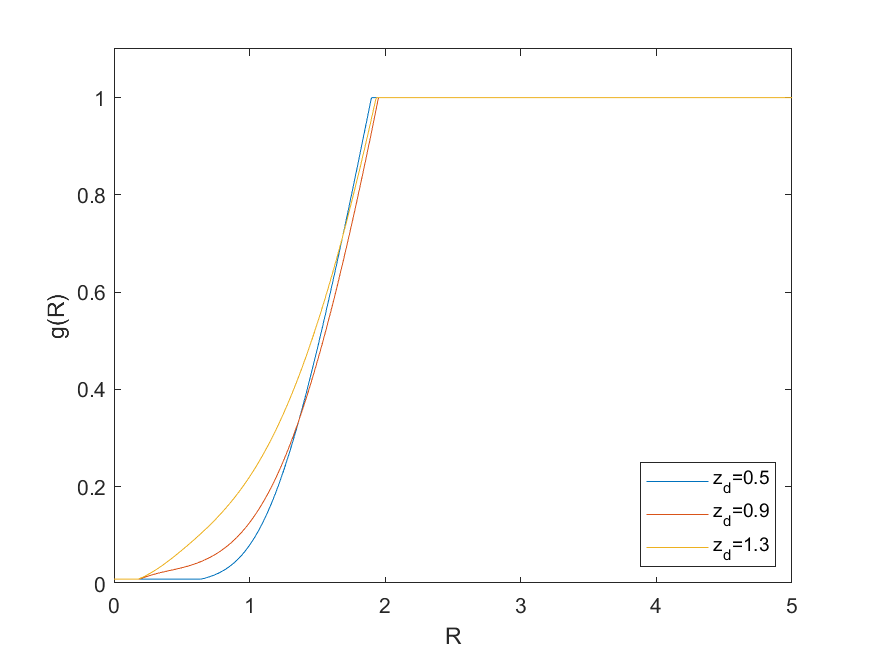}
   \caption{Surface deformation (left), and boundary data (right), plotted for $\gamma=500$, $\Omega=0.5$, $z_d=1.3, 0.9, 0.5$.}
  \label{fig:sd_results6}
\end{figure}

\begin{figure}
   \centering
   \includegraphics[width=0.49\linewidth]{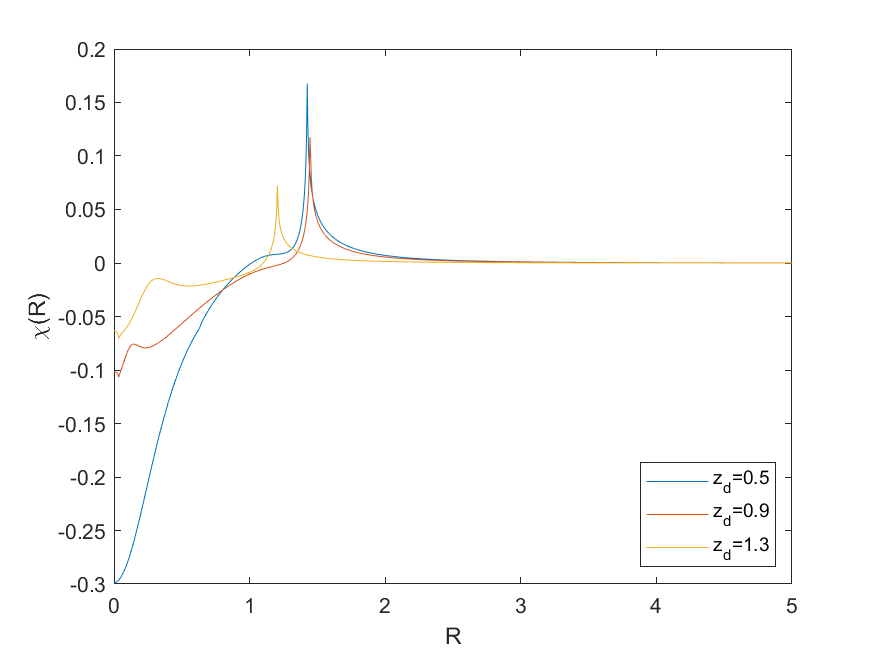}
   \includegraphics[width=0.49\linewidth]{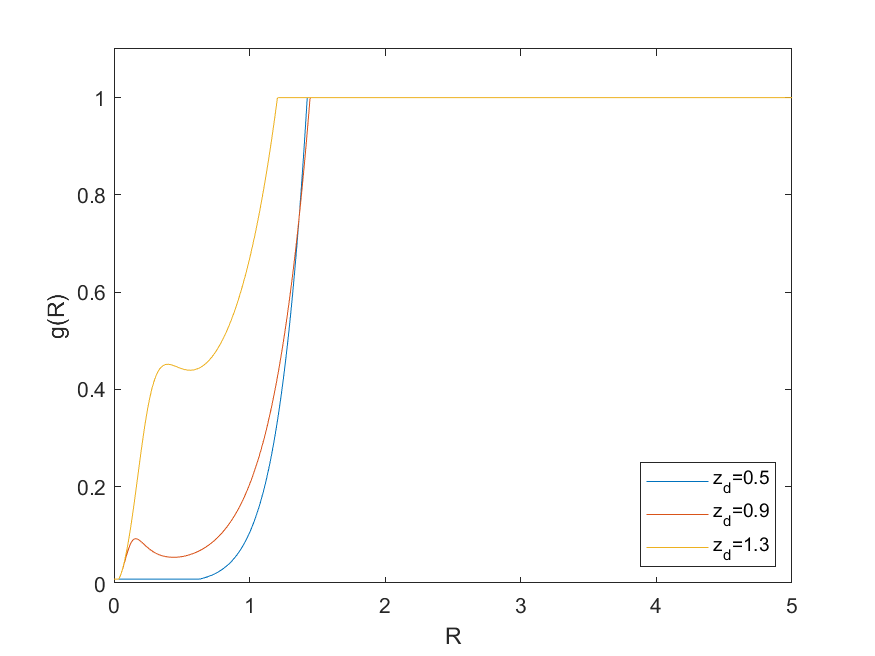}
   \caption{Surface deformation (left), and boundary data (right), plotted for $\gamma=500$, $\Omega=0.1$, $z_d=1.3, 0.9, 0.5$.}
  \label{fig:sd_results7}
\end{figure}

\begin{figure}
   \centering
   \includegraphics[width=0.49\linewidth]{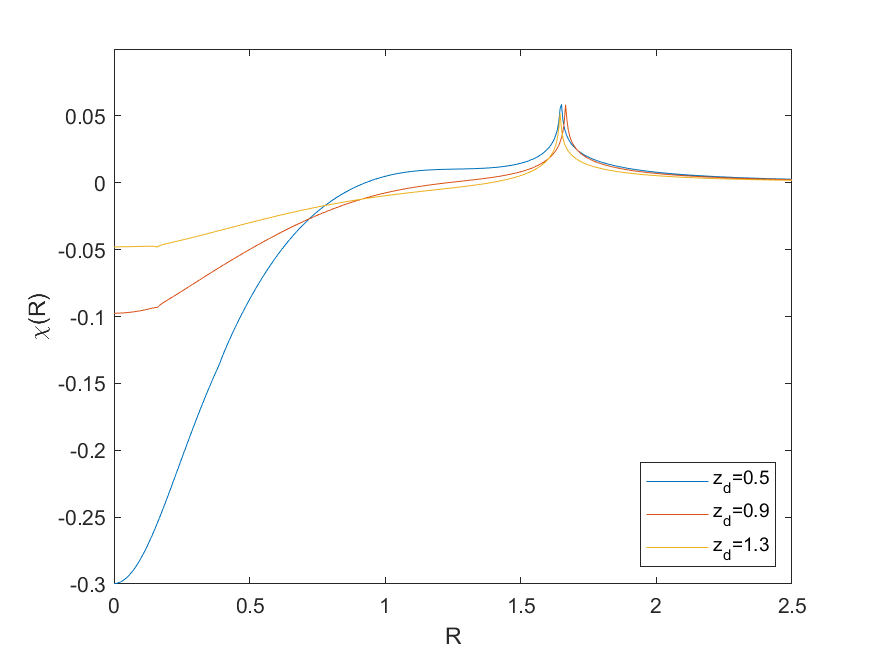}
   \includegraphics[width=0.49\linewidth]{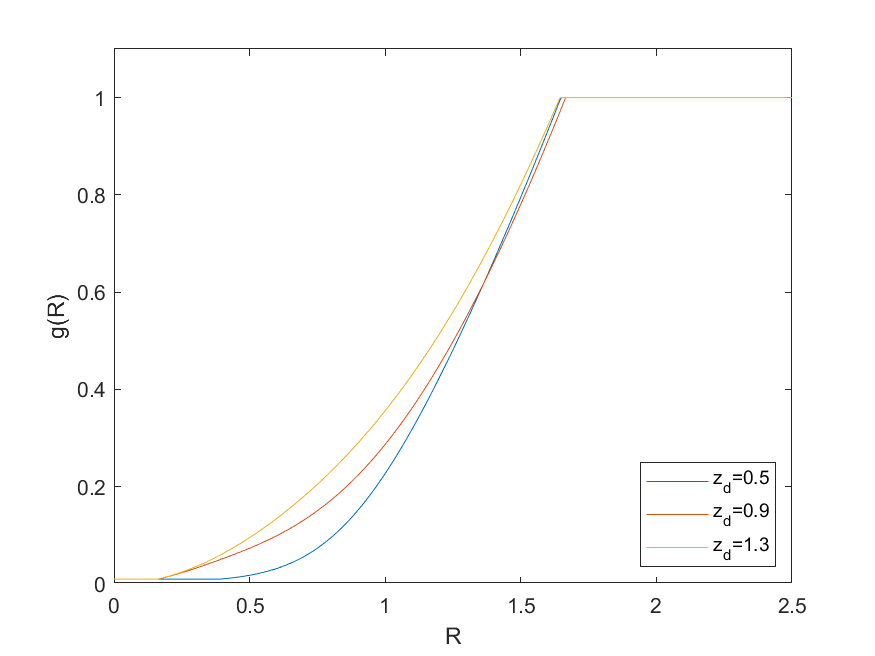}
   \caption{Surface deformation (left), and boundary data (right), plotted for $\gamma=100$, $\Omega=1.0$, $z_d=1.3, 0.9, 0.5$.}
  \label{fig:sd_results8}
\end{figure}

\begin{figure}
   \centering
   \includegraphics[width=0.49\linewidth]{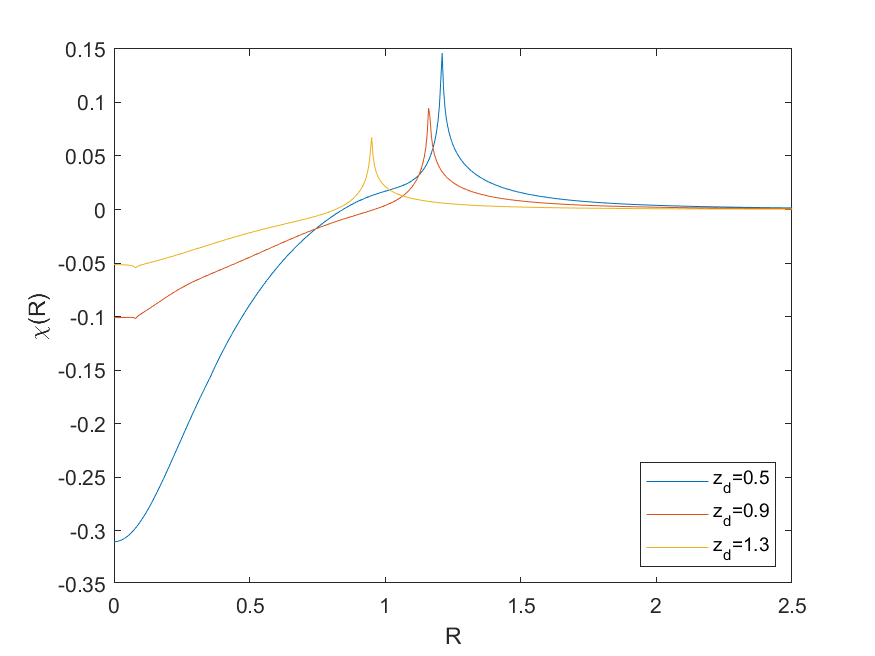}
   \includegraphics[width=0.49\linewidth]{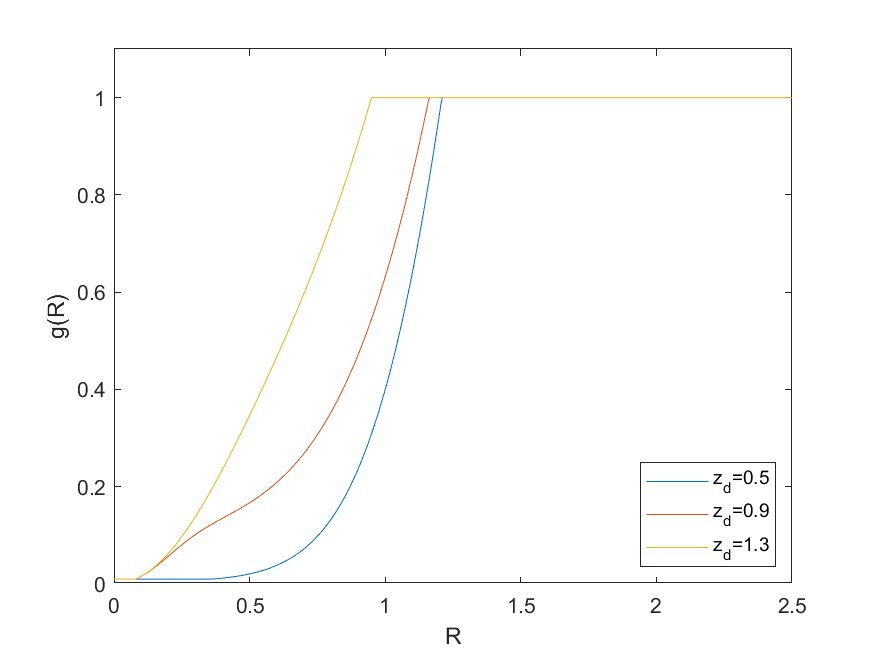}
   \caption{Surface deformation (left), and boundary data (right), plotted for $\gamma=100$, $\Omega=0.5$, $z_d=1.3, 0.9, 0.5$.}
  \label{fig:sd_results9}
\end{figure}

\begin{figure}
   \centering
   \includegraphics[width=0.49\linewidth]{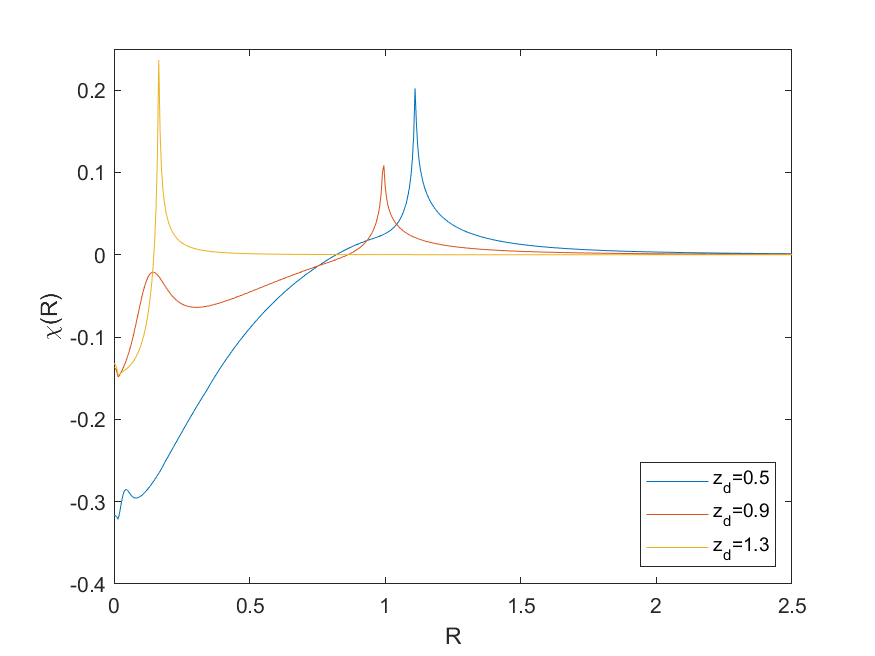}
   \includegraphics[width=0.49\linewidth]{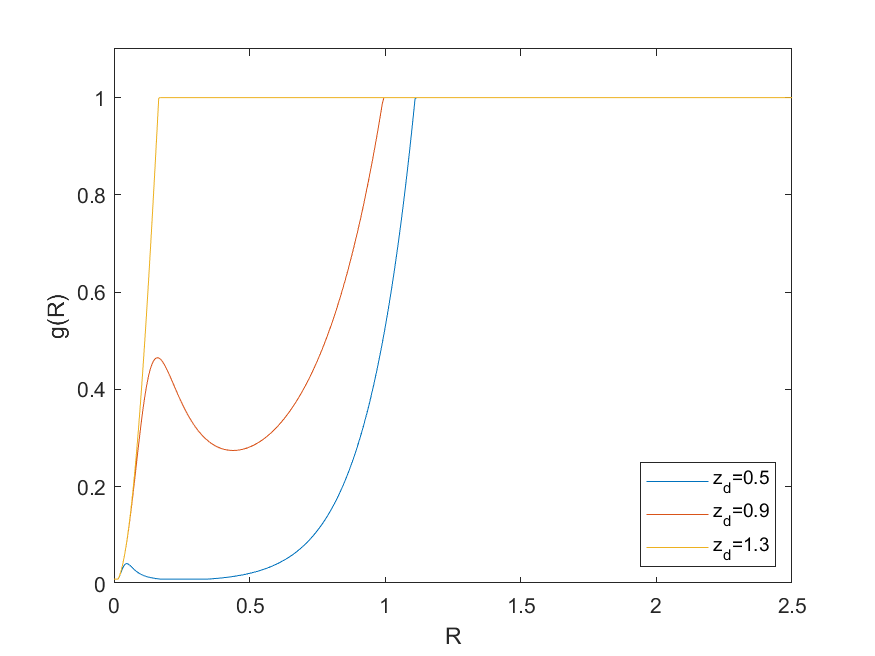}
   \caption{Surface deformation (left), and boundary data (right), plotted for $\gamma=100$, $\Omega=0.1$, $z_d=1.3, 0.9, 0.5$.}
  \label{fig:sd_results10}
\end{figure}

We now follow the above procedure  to compute surface deformation profiles, for each combination of parameter values $\gamma=1000, 500, 100$, $\Omega=1,0.5,0.1$ and $z_d=1.3, 0.9, 0.5$, with the respective profiles being shown on the left of figures~\ref{fig:sd_results2}--\ref{fig:sd_results10} (note that the horizontal and vertical scales vary between plots).  On the right of each of these figures, we also plot the corresponding boundary data $g$, as given by~(\ref{eqn:h25}).  The profile in figure~\ref{fig:sd_results6}
for $z_d=0.9$ corresponds to the voidage field seen in figure~\ref{fig:opacity}.  The plots of voidage field $E$ for each of the parameter combinations shown in figures~\ref{fig:sd_results2}--\ref{fig:sd_results10} can be found in \cite[figures~9--11]{NL24}, but broadly speaking each is qualitatively (though not quantitatively) comparable to that shown in figure~\ref{fig:opacity}.

In each of figures~\ref{fig:sd_results2}--\ref{fig:sd_results10} we see qualitatively comparable features in the computed profiles.  For each case, $\chi(R)<0$ near $R=0$, with a deeper dip (i.e., larger $|\chi(R)|$) when $z_d$ is smaller;  this can be interpreted as a deeper hollow in the surface immediately underneath the hovering helicopter as $z_d$ decreases.  As $R$ increases, we see $\chi(R)$ increase, with a cusp being seen in each case - this appears to coincide with the value of $R$ for which the gradient of the boundary data $g$ has a discontinuity. We also note that when $\Omega=0.1$ (figures~\ref{fig:sd_results4}, \ref{fig:sd_results7} and~\ref{fig:sd_results10}) oscillations in $\chi(R)$ for small values of $R$ are present. As noted in~\cite{NL24}, for certain parameter combinations (such as that seen in figure~\ref{fig:sd_results10}) there is a qualitative change in the features seen in the boundary data, as shown in~\cite[figure~12]{NL24}, and the results seen here are consistent with that.

\begin{figure}
   \centering
   \includegraphics[width=0.49\linewidth]{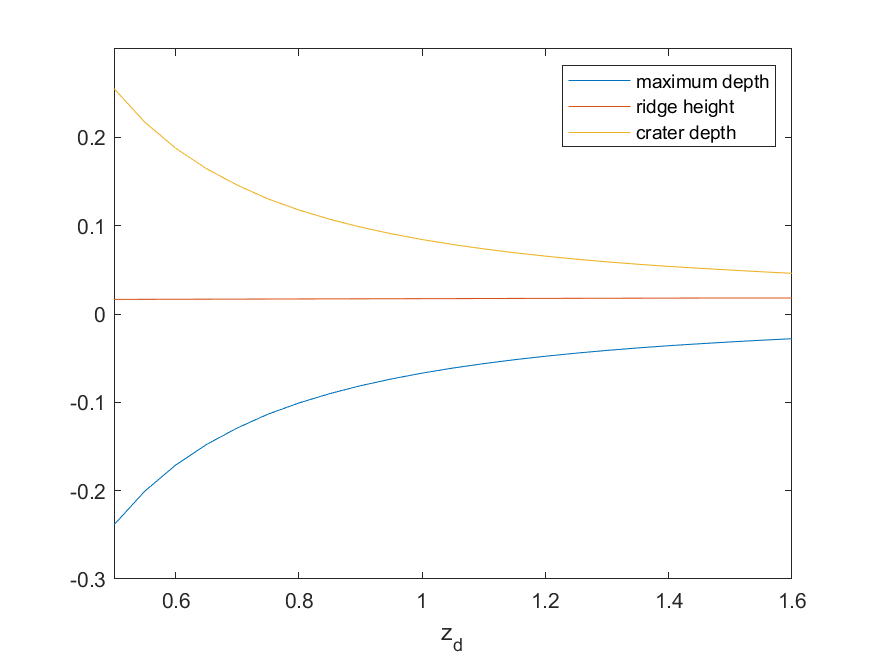}
   \includegraphics[width=0.49\linewidth]{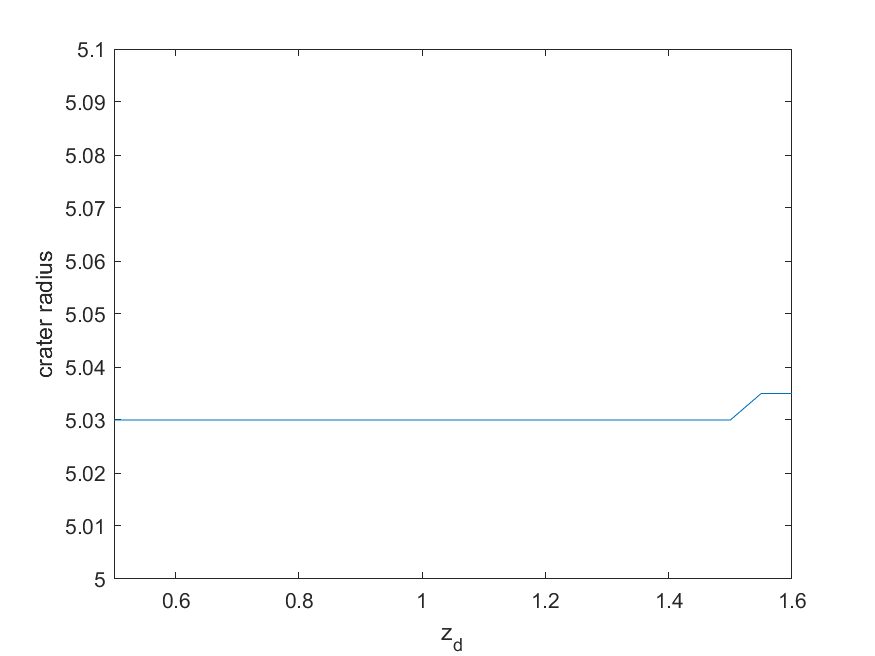}
   \caption{Surface deformation features, plotted for $\gamma=1000$, $\Omega=1.0$, $z_d\in[0.5,1.6]$.}
  \label{fig:sd_crater2}
\end{figure}

\begin{figure}
   \centering
   \includegraphics[width=0.49\linewidth]{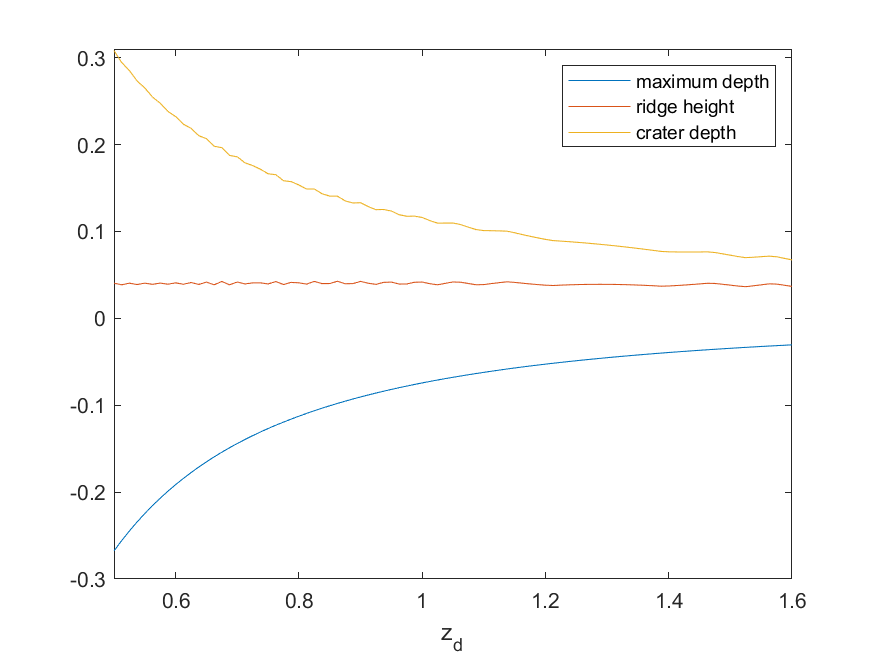}
   \includegraphics[width=0.49\linewidth]{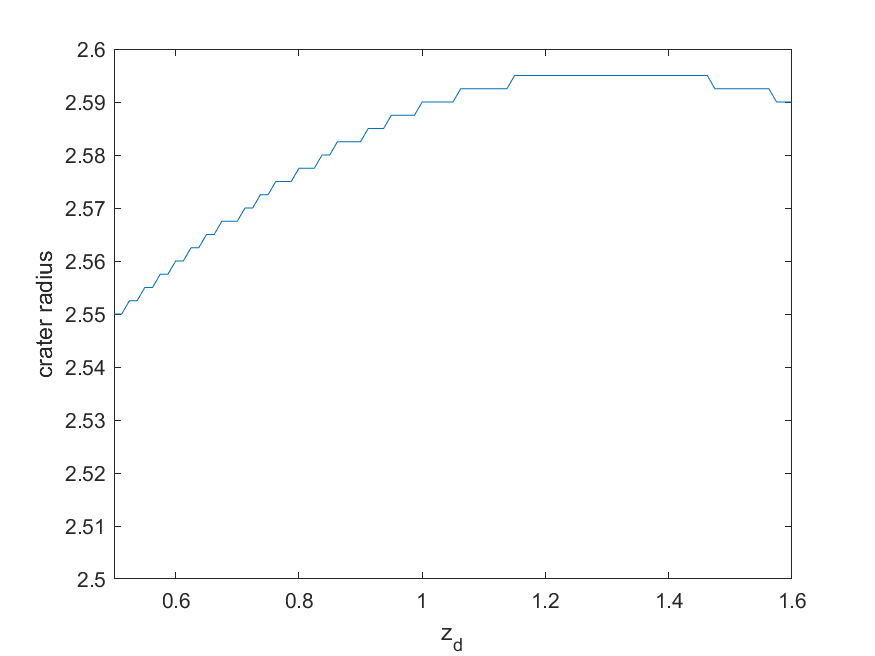}
   \caption{Surface deformation features, plotted for $\gamma=1000$, $\Omega=0.5$, $z_d\in[0.5,1.6]$.}
  \label{fig:sd_crater3}
\end{figure}

\begin{figure}
   \centering
   \includegraphics[width=0.49\linewidth]{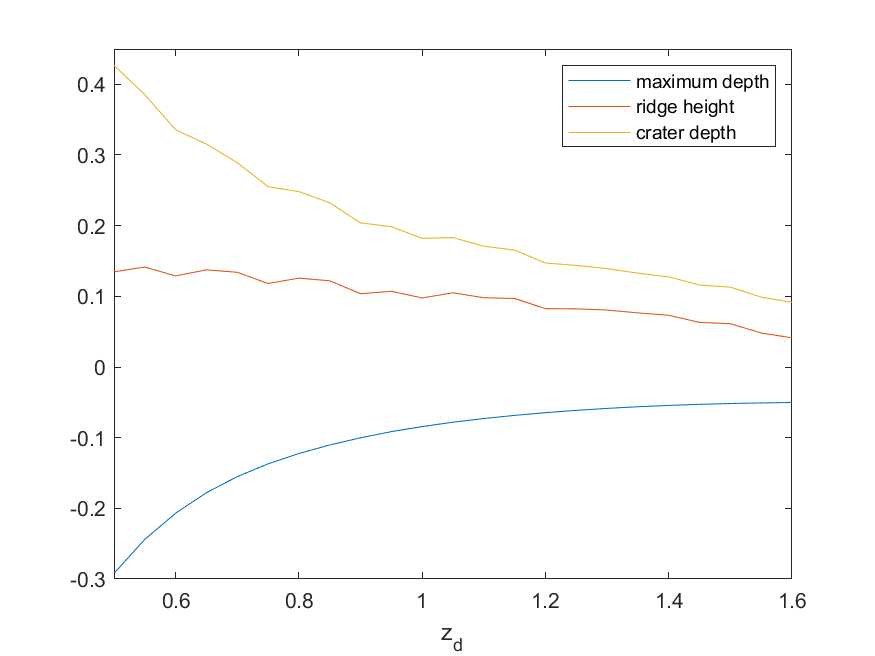}
    \includegraphics[width=0.49\linewidth]{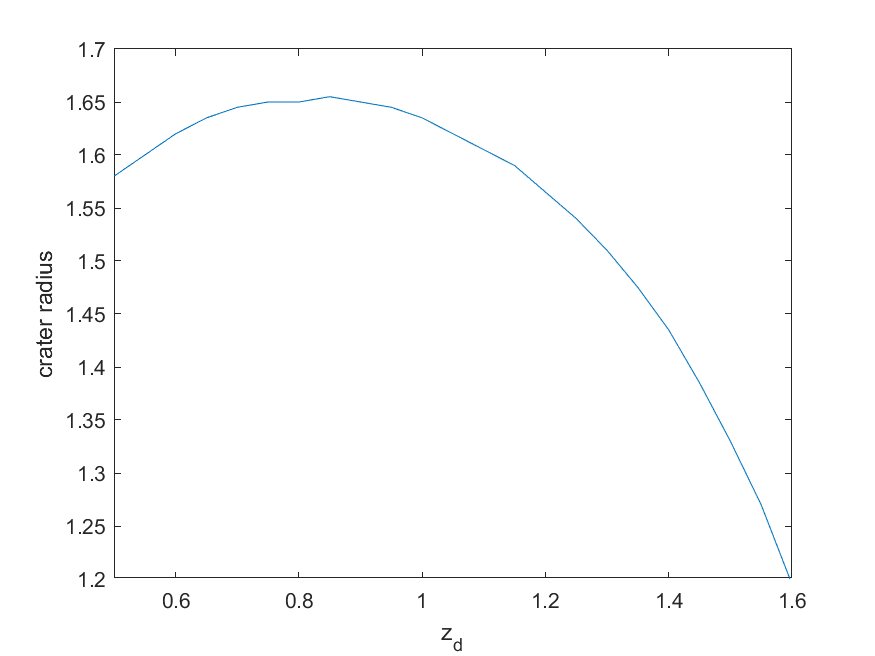}
   \caption{Surface deformation features, plotted for $\gamma=1000$, $\Omega=0.1$, $z_d\in[0.5,1.6]$.}
  \label{fig:sd_crater4}
\end{figure}

\begin{figure}
   \centering
   \includegraphics[width=0.49\linewidth]{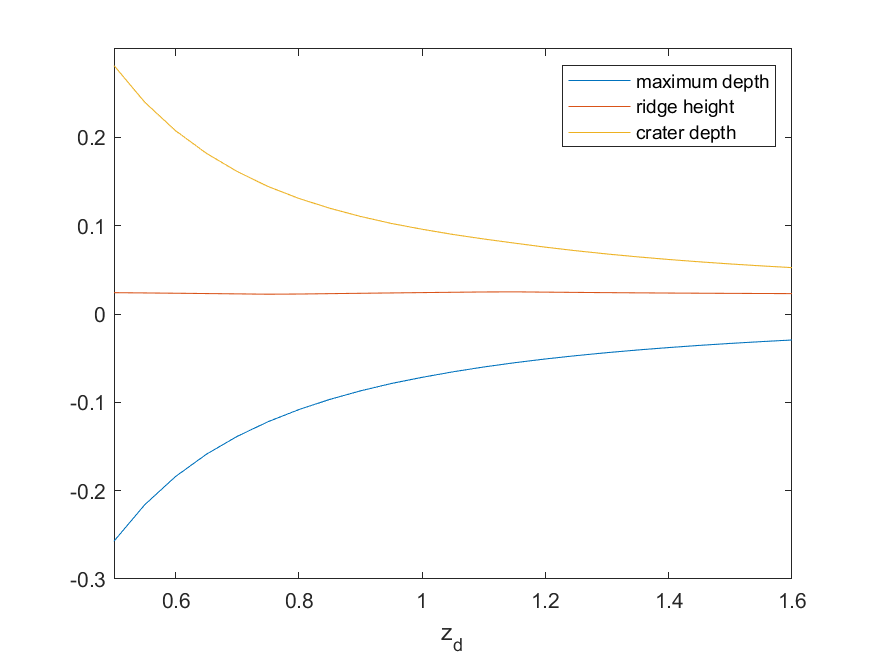}
   \includegraphics[width=0.49\linewidth]{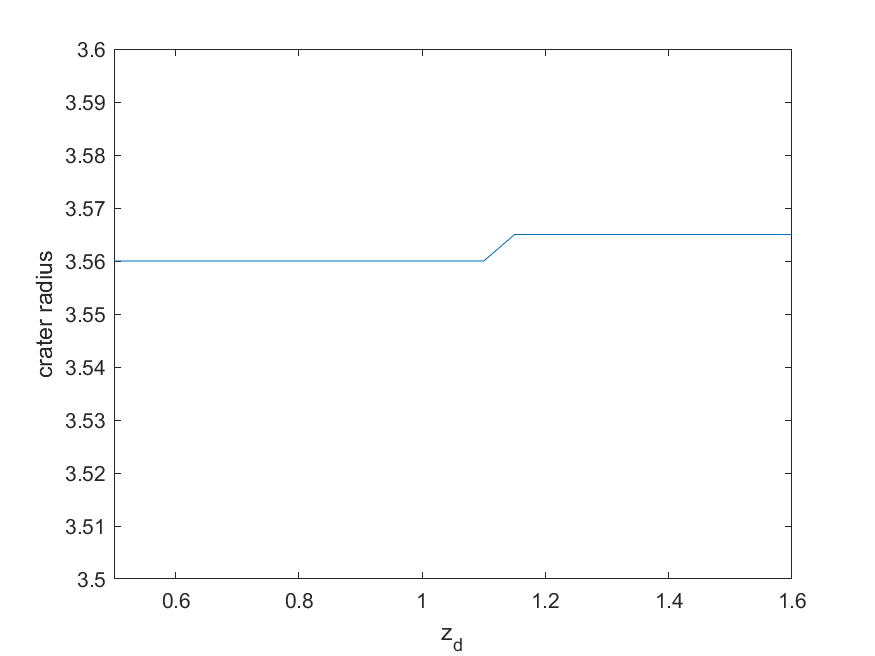}
   \caption{Surface deformation features, plotted for $\gamma=500$, $\Omega=1.0$, $z_d\in[0.5,1.6]$.}
  \label{fig:sd_crater5}
\end{figure}

\begin{figure}
   \centering
   \includegraphics[width=0.49\linewidth]{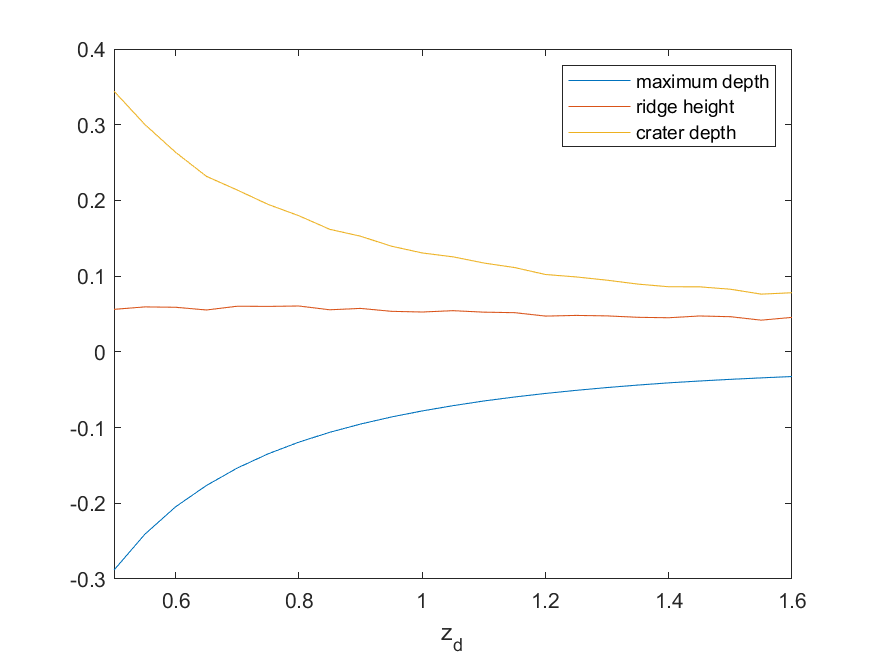}
   \includegraphics[width=0.49\linewidth]{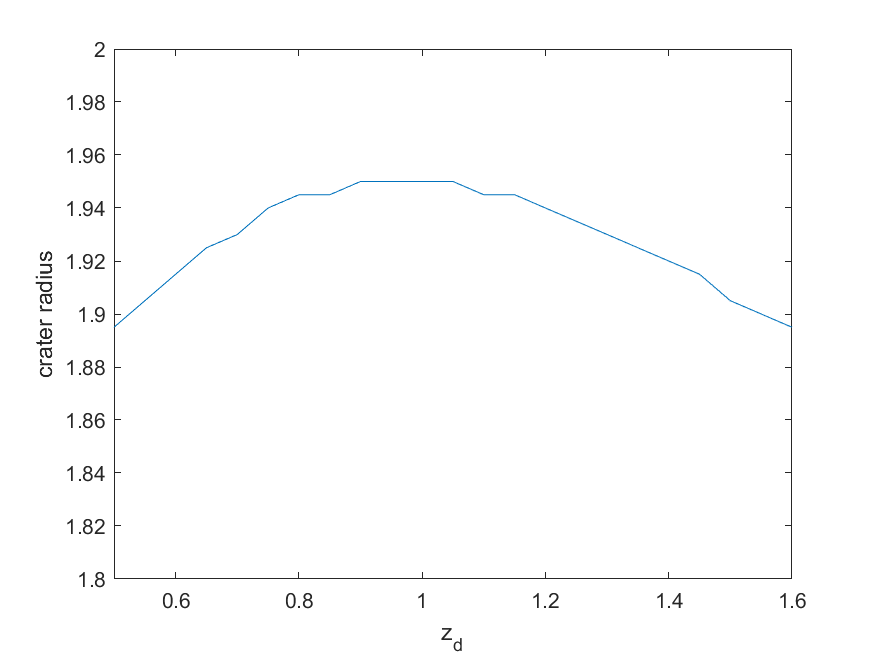}
   \caption{Surface deformation features, plotted for $\gamma=500$, $\Omega=0.5$, $z_d\in[0.5,1.6]$.}
  \label{fig:sd_crater6}
\end{figure}

\begin{figure}
   \centering
   \includegraphics[width=0.49\linewidth]{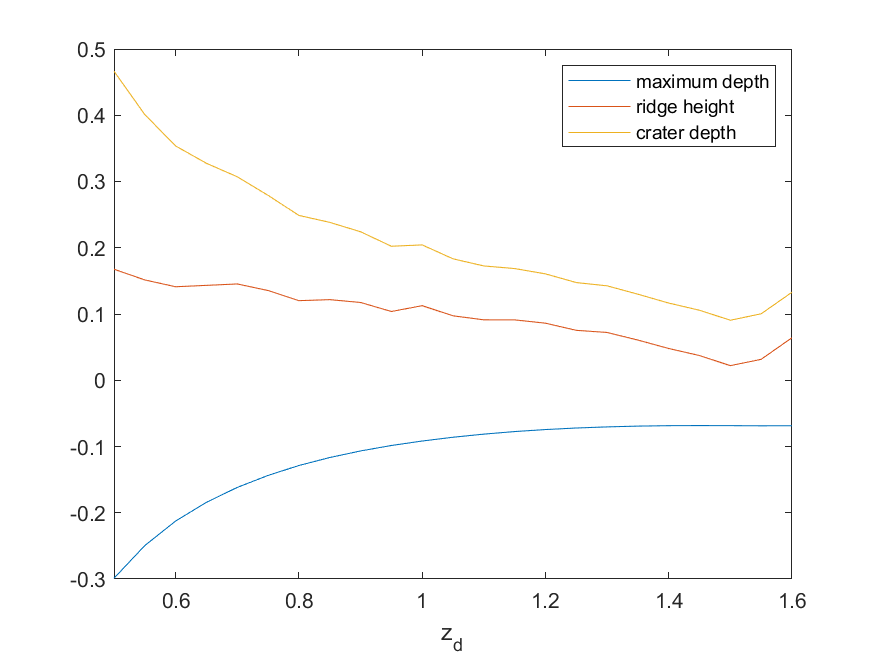}
   \includegraphics[width=0.49\linewidth]{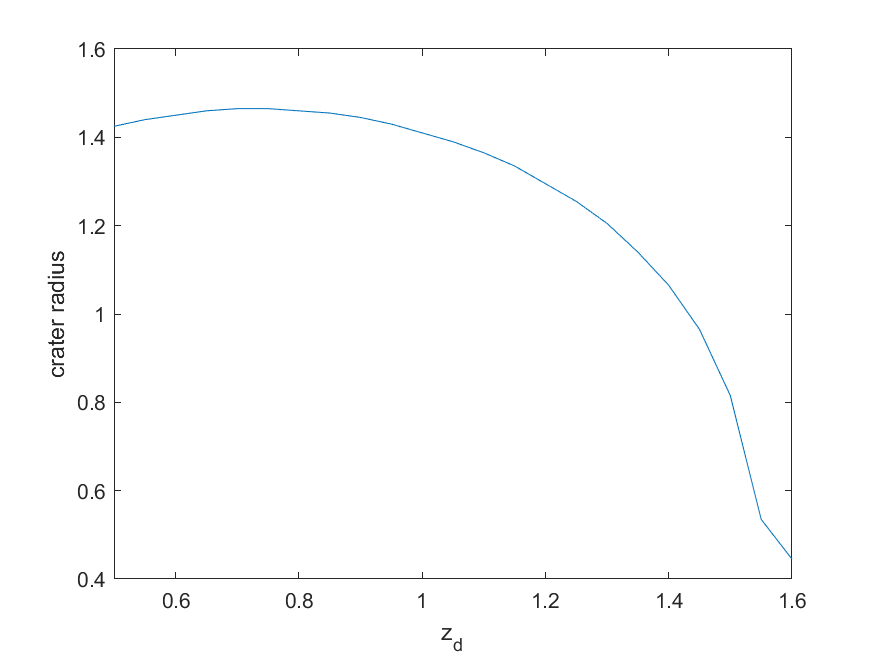}
   \caption{Surface deformation features, plotted for $\gamma=500$, $\Omega=0.1$, $z_d\in[0.5,1.6]$.}
  \label{fig:sd_crater7}
\end{figure}

\begin{figure}
   \centering
   \includegraphics[width=0.49\linewidth]{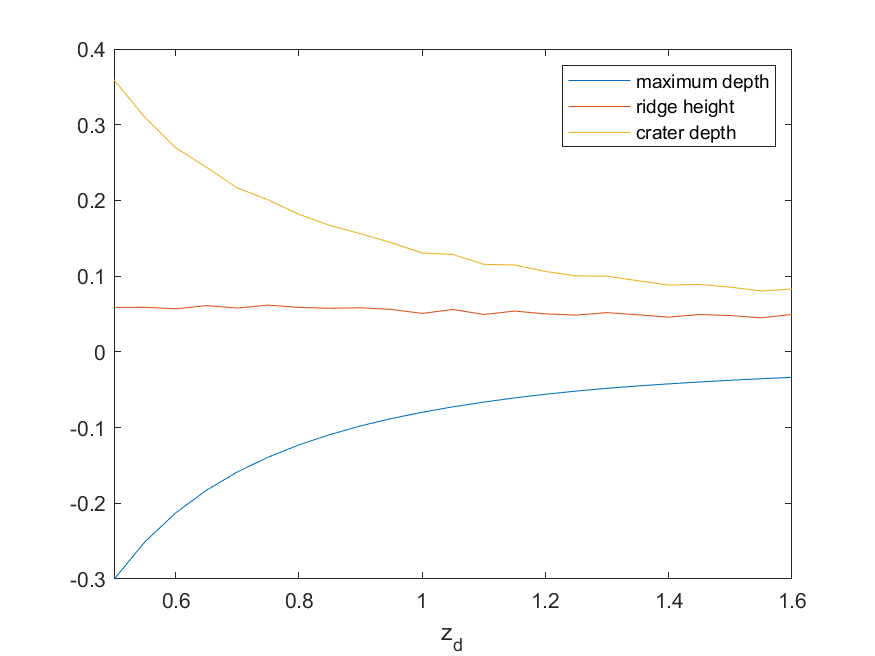}
   \includegraphics[width=0.49\linewidth]{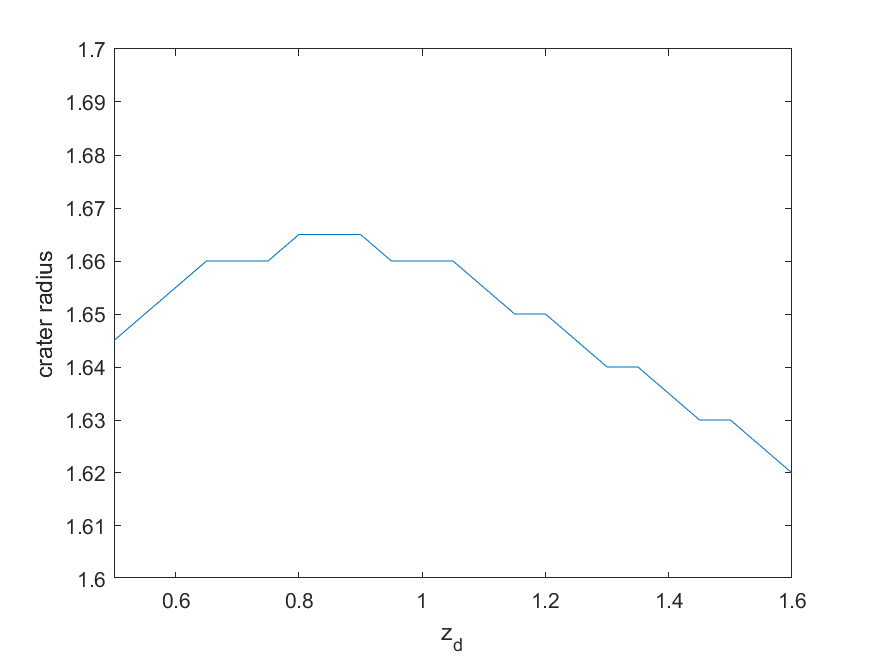}
   \caption{Surface deformation features, plotted for $\gamma=100$, $\Omega=1.0$, $z_d\in[0.5,1.6]$.}
  \label{fig:sd_crater8}
\end{figure}

\begin{figure}
   \centering
   \includegraphics[width=0.49\linewidth]{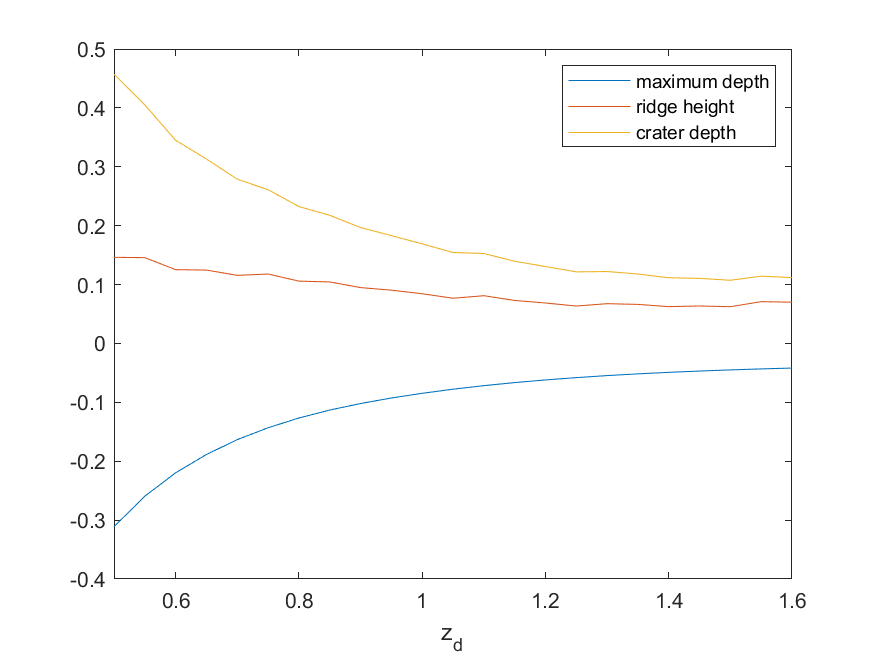}
   \includegraphics[width=0.49\linewidth]{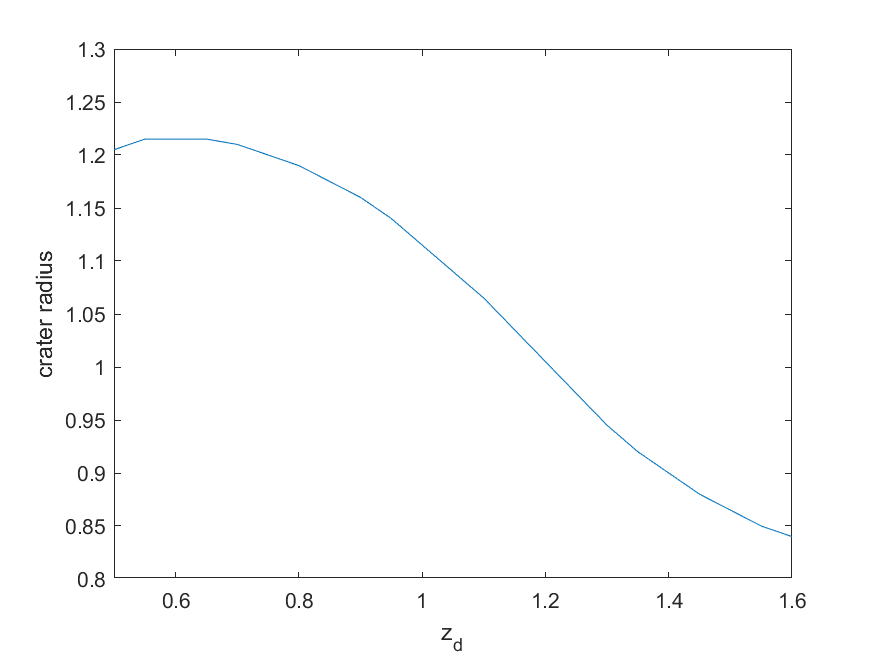}
   \caption{Surface deformation features, plotted for $\gamma=100$, $\Omega=0.5$, $z_d\in[0.5,1.6]$.}
  \label{fig:sd_crater9}
\end{figure}

\begin{figure}
   \centering
   \includegraphics[width=0.49\linewidth]{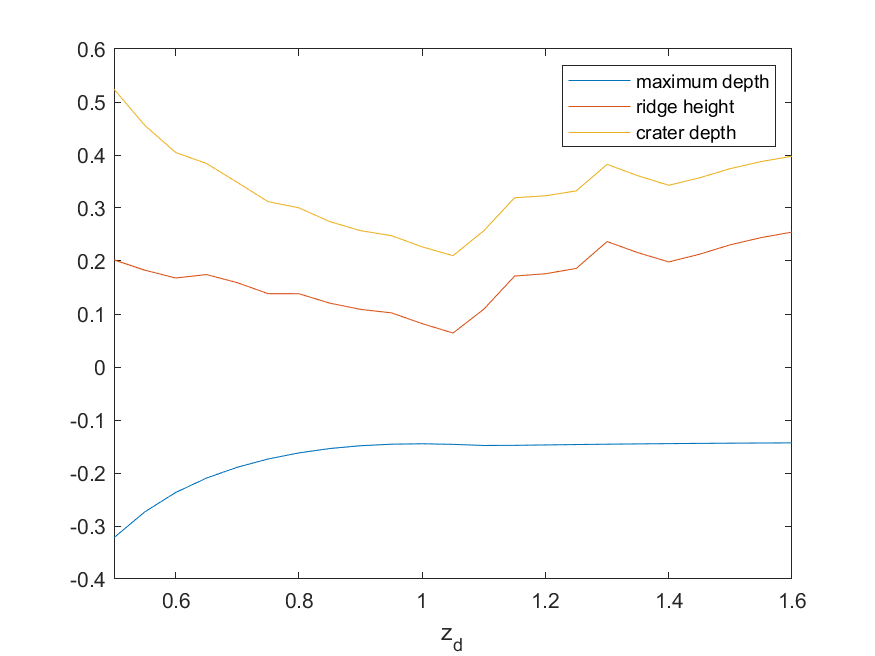}
   \includegraphics[width=0.49\linewidth]{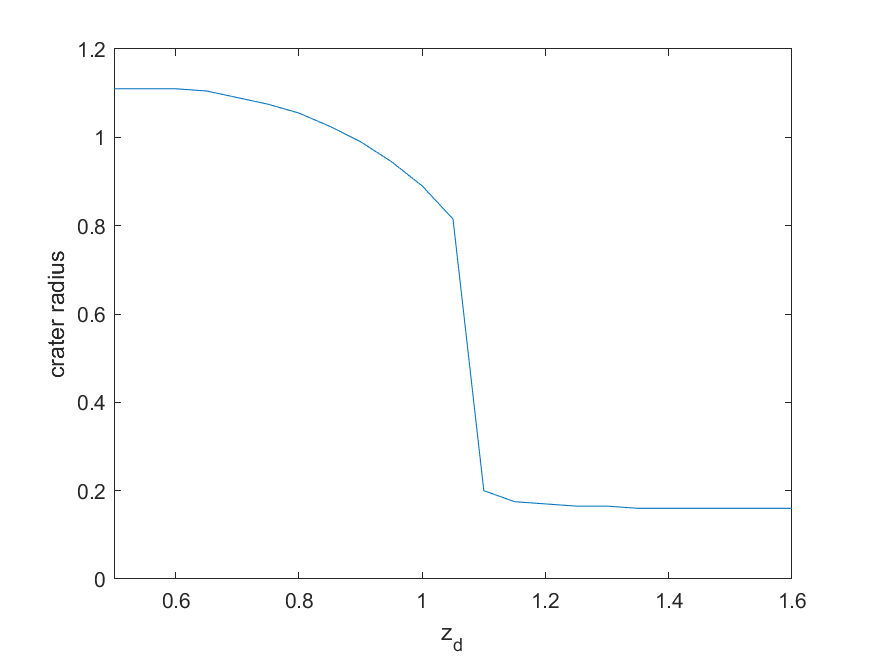}
   \caption{Surface deformation features, plotted for $\gamma=100$, $\Omega=0.1$, $z_d\in[0.5,1.6]$.}
  \label{fig:sd_crater10}
\end{figure}

To explore this further, for each combination of parameter values $\gamma=1000, 500, 100$, $\Omega=1,0.5,0.1$, we consider a wider range of values of $z_d\in[0.5,1.6]$, and for each of the nine combinations of $\gamma$ and $\Omega$ we plot, in figures~\ref{fig:sd_crater2}--\ref{fig:sd_crater10}: the maximum depression depth, i.e., the value of $\chi(0)$, the ridge peak height, i.e., the maximum value of $\chi(R)$, and the crater depth, i.e., $\max(\chi(R))-\chi(0)$, as well as the crater radius, i.e., the value of $R$ at which $\chi(R)$ attains its maximum, each against $z_d$.  Note that the horizontal and vertical scales vary between plots.

For all parameter combinations except for $\gamma=100$, $\Omega=0.1$ (i.e., for figures~\ref{fig:sd_crater2}--\ref{fig:sd_crater9}), the maximum depression depth is generally greater in magnitude when $z_d$ is smaller (i.e., a deeper depression the closer to the ground that the helicopter is hovering).  The ridge height either decreases or remains constant as $z_d$ increases in most cases, with the case for $\gamma=100$, $\Omega=0.1$, as seen in figure~\ref{fig:sd_crater10}, again a notable exception.  The small oscillations in the ridge height, as a function of $z_d$, appear to be at least in part a facet of our numerical approximation scheme, noting the sensitivity in attempting to measure numerically the height of the cusp representing the ridge height.  In most cases, the variation in ridge height as $z_d$ varies is less than the variation in depression depth, and hence in most cases the crater depth also decreases as $z_d$ increases, again with the exception of figure~\ref{fig:sd_crater10}, which we will return to shortly.  The crater radius is generally either approximately equal for all values of $z_d$ (as in figures~\ref{fig:sd_crater2}, \ref{fig:sd_crater5}), or else increasing and then decreasing as $z_d$ increases (as in figures~\ref{fig:sd_crater3}, \ref{fig:sd_crater4}, \ref{fig:sd_crater6}--\ref{fig:sd_crater9}).

For the case $\gamma=100$, $\Omega=0.1$, as shown in figure~\ref{fig:sd_crater10}, the ridge height decreases as $z_d$ increases, up to $z_d\approx1.1$, before increasing, and the same behaviour is seen for the overall crater depth (i.e., decrease then increase as $z_d$ increases).  Furthermore, the crater radius decreases slowly as $z_d$ increases from $z_d=0.5$, before decreasing very rapidly around $z_d\approx1.1$, and then levelling off.

\begin{figure}
   \centering
   \includegraphics[width=0.49\linewidth]{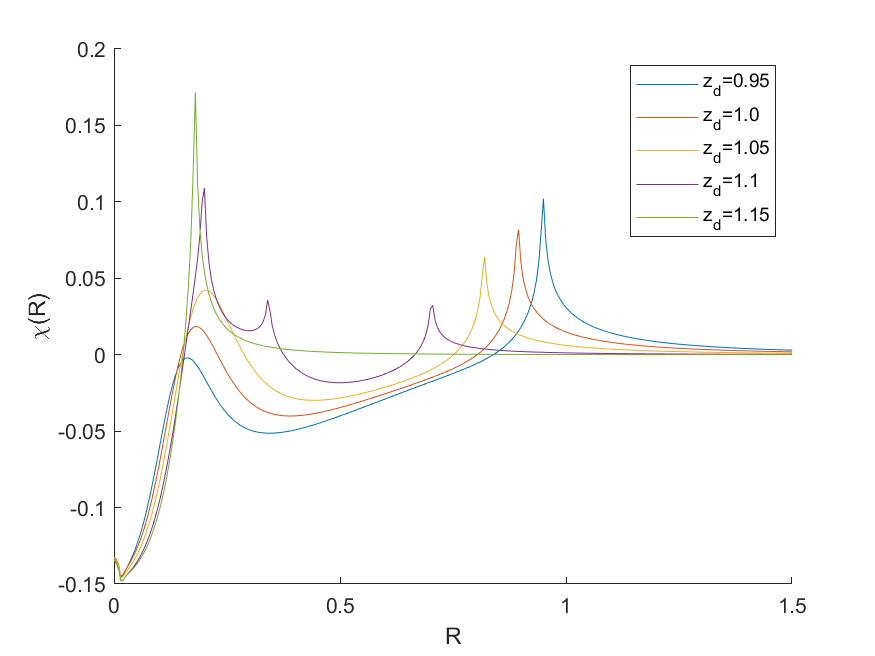}
   \includegraphics[width=0.49\linewidth]{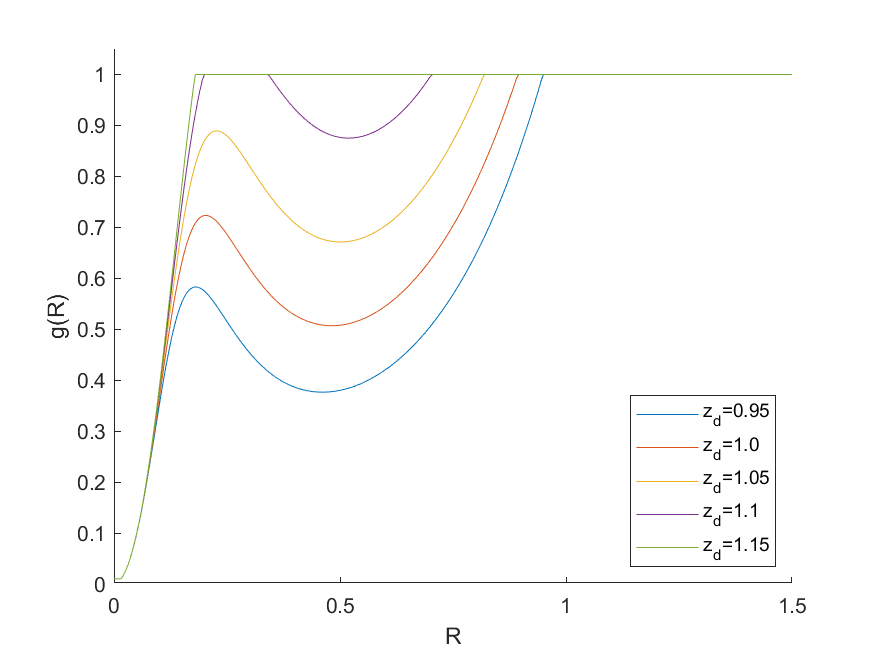}
   \caption{Surface deformation (left), and boundary data (right), plotted for $\gamma=100$, $\Omega=0.1$, $z_d\in[0.95,1.15]$.}
  \label{fig:sd_results10_zdrange}
\end{figure}

To explore this behaviour further, in figure~\ref{fig:sd_results10_zdrange} we plot the surface deformation and the boundary data over a range of values of $z_d\in[0.95,1.15]$ (compare to figure~\ref{fig:sd_results10}, where the comparable results were plotted for $z_d=0.5,0.9,1.3$).
On the left of figure~\ref{fig:sd_results10_zdrange}, we see that as $z_d$ increases from $z_d=0.95$, the ridge height decreases and the value of $R$ at which $\chi(R)$ is maximised decreases.  At the same time, another ridge starts to form at a lower value of $R$, and at a critical crossover value (between $z_d=1.05$ and $z_d=1.1$), the height of the ``inner ridge'' becomes larger than the height of the ``outer ridge''.  For $z_d=1.1$, we see three such peaks, but as $z_d$ increases further to $z_d=1.15$, two peaks flatten out and only one remains.  The reason for this behaviour becomes clearer on examining the boundary data, on the right of figure~\ref{fig:sd_results10_zdrange}, where we see that for $z_d=1.1$ there are two separate regions where $g(R)=1$, with these two regions having merged together for $z_d=1.15$.

\section{Conclusion}
\label{sec:conc}
In this paper, we have described the application of a recently derived mathematical model for wind-generated particle fluid flow fields to the helicopter cloud problem.  We have introduced a measure for the light transmission defect from the far field to the helicopter cab, as a function of the observation angle, and have demonstrated how visibility depends on the relationship between the height of the helicopter, the ratio of swirl velocity to downdraft velocity, and the balance between lift and gravity in the interfacial layer between the sand bed and the fluidized region.  In particular, for certain parameter choices we see a potentially counterintuitive result suggesting that pilot visibility may be improved in some viewing directions if the helicopter were hovering at a lower altitude.  We have also computed the deformation of the upper surface of the particle bed, and have shown how key features of the deformation, including crater depth and ridge height, may be sensitive to variations in key parameters.  In future work, it would be interesting to undertake experimental work in order to further validate our model.

\bibliography{helicopter_bib}

\end{document}